\documentclass[a4paper,fleqn]{cas-sc}

\usepackage[authoryear]{natbib}

\usepackage{subcaption}

\begin{document}
\let\WriteBookmarks\relax
\def\floatpagepagefraction{1}
\def\textpagefraction{.001}

\shorttitle{TBD}

\shortauthors{Zahura et~al.}

\title [mode = title]{Impact of Topography and Climate on Post-fire Vegetation Recovery Across Different Burn Severity and Land Cover Types through Machine Learning}    

\author[1]{Faria Tuz Zahura}[orcid=0000-0001-9005-9951]
\credit{Conceptualization, Data curation, Formal analysis, Methodology, Investigation, Software, Roles/Writing - original draft}

\author[1]{Gautam Bisht}[orcid=0000-0001-6641-7595]
\credit{Conceptualization, Supervision, Writing - review \& editing}

\author[1]{Zhi Li}[orcid=0000-0002-6240-8971]
\credit{Resources, Writing - review \& editing}

\author[2]{Sarah McKnight}[orcid=0000-0002-6013-193X]
\credit{Methodology, Writing - review \& editing}

\author[1]{Xingyuan Chen}[orcid=0000-0003-1928-5555]
\credit{Conceptualization,  Funding acquisition, Project administration, Supervision, Writing - review \& editing}
\cormark[1]
\cortext[cor1]{Corresponding author}
\ead{xingyuan.chen@pnnl.gov}

\address[1]{Atmospheric, Climate, \& Earth Sciences Division, Pacific Northwest National Laboratory, Richland, WA, United States}

\address[2]{Department of Geology and Environmental Geosciences, University of Dayton, Dayton, OH, United States}

\begin{abstract}
Wildfires significantly disturb ecosystems by altering forest structure, vegetation ecophysiology, and soil properties. Understanding the complex interactions between topographic and climatic conditions in post-wildfire recovery is crucial. This study investigates the interplay between topography, climate, burn severity, and years after fire on vegetation recovery across dominant land cover types (evergreen forest, shrubs, and grassland) in the Pacific Northwest region. Using Moderate Resolution Imaging Spectroradiometer (MODIS) data, we estimated vegetation recovery by calculating the incremental Enhanced Vegetation Index change during post-fire years. A machine learning technique, random forest (RF), was employed to map relationships between the input features (elevation, slope, aspect, precipitation, temperature, burn severity, and years after fire) and the target (incremental EVI recovery) for each land cover type. Variable importance analysis and partial dependence plots were generated to understand the influence of individual features. The observed and predicted incremental EVI values showed good matches, with $R^2$ values of 0.99 for training and 0.89-0.945 for testing. The study found that climate variables, specifically precipitation and temperature, were the most important features overall, while elevation played the most significant role among the topographic factors. Partial dependence plots revealed that lower precipitation tended to cause a reduction in vegetation recovery for varying temperature ranges across land cover types. These findings can aid in developing targeted strategies for post-wildfire forest management, considering the varying responses of different land cover types to topographic, climatic, and burn severity factors.
\end{abstract}



\begin{highlights}

\item Factors impacting post-fire vegetation recovery were analyzed with machine learning.
\item Climate factors had the most influence in controlling the recovery process.
\item Topographic controls were mostly evident in evergreen forests.

\end{highlights}


\maketitle

\section{Introduction}
Climate change-induced fuel dryness has led to a rise in wildfire occurrences within the forests of the western United States \citep{abatzoglou2016impact, gergel2017effects}. Wildfires have a profound impact on watersheds, as they modify soil properties, alter vegetation composition, disrupt hydrologic cycles, result in biomass destruction, and contribute to a decrease in water quality \citep{fultz2016forest, hrelja2020wildfire, chen2022review, robinne2020wildfire, kraaij2018assessment, taylor2021controls}. While the ecosystem recovery from fire process is a complex process that depends on the post-fire climate, unique characteristics of the fire event, and the watershed characteristics \citep{nelson2013influence, ireland2015exploring, Maia2014}, it is vital to understand how vegetation recovers after fires to make informed decisions regarding land management and restoration planning. \citep{Casady2010, vallejo2012post}.

Previous research has identified soil moisture availability as the primary factor that influences the recovery process \citep{wilson2015climatic, viana2020identifying, lippok2013forest, johnstone2010sensitive}. This moisture availability, in turn, is dependent on several climatic and topographic conditions. Increasing post-fire precipitation has been linked with a higher vegetation recovery rate \citep{vo2020remote, kinoshita2011spatial, chen2022remote}. Conversely, warmer temperatures and reduced precipitation promote drought conditions, which hinder recovery by limiting vegetation recruitment \citep{wilson2015climatic, chappell1996fire, sanchez2006performance, harvey2016high, haffey2018limits}. Climate change is anticipated to exacerbate drought conditions across the western United States, which may adversely impact post-fire vegetation recovery \citep{cook2018climate, peltier2019legacies}. Besides climatic factors, topography also impacts local microclimate, and consequently on soil moisture availability \citep{ireland2015exploring}. Elevation, which shapes local temperature and precipitation, can exert a positive or negative influence on the vegetation recovery \citep{viana2020identifying, meng2015effects, viana2017assessment, daskalakou1996aleppo}. Moreover, terrains with steeper slope tend to have low vegetation recovery due to their smaller water holding capacity and higher soil erosion \citep{Evangelides2020, pereira2016short}. The orientation of the slope, called the aspect, may also play a role in ecosystem recovery because it influences the solar exposure of the landscape. For example, the equatorial facing slopes that receive higher radiation are found to have elevated temperatures and dryer conditions \citep{griffiths2009effects}. Consequently, these slopes may undergo a slower rate of post-fire recovery compared to polar facing slopes with a higher moisture content \citep{ireland2015exploring, viana2017assessment, kinoshita2011spatial, petropoulos2014quantifying, rengers2020landslides}. 

Remote sensing data has been extensively employed to examine the ecological changes post a fire due to its cost-effectiveness and greater spatial coverage compared to labor-intensive and spatially restricted field surveys \citep{fox2008using, lentile2006remote}. As a means of assessing vegetation recovery post fires, researchers have widely adopted the normalised difference vegetation index (NDVI) as a proxy to vegetation biomass \citep{ireland2015exploring, leon2012using, diaz2002satellite, goetz2006using, hao2022long}. NDVI is calculated using light reflected in the near-infrared (NIR) and red (RED) spectrum as (NIR-RED)/(NIR+RED) \citep{rouse1974monitoring}. Healthy vegetation reflects more NIR light while absorbing significant portion of RED light for photosynthesis. Conversely, sparse and unhealthy vegetation exhibits a smaller disparity between reflected NIR and RED light. Thus, NDVI can be used to identify immediate post-fire vegetation destruction and gradual rejuvenation over time \citep{diaz2002satellite, Evangelides2020, prodon2021assessing}. However, the surrounding vegetation background and atmospheric conditions can influence the accuracy and reliability of NDVI \citep{wittenberg2007spatial, huete1985spectral, holben1986characteristics}. The atmospheric aerosols scatter light based on its wavelength and the effect on the red portion of the spectrum is much larger than on the NIR spectrum. To improve the quality of NDVI, researchers adopted the enhanced vegetation index (EVI) by using the difference in the blue and red reflectances to correct for atmospheric influences and also by adjusting for canopy background \citep{liu1995feedback, gao2000optical,chen2005monitoring}. EVI has been used as a robust proxy to inform vegetation recovery following disturbances \citep{wittenberg2007spatial, kinoshita2011spatial, chen2005monitoring, wang2022vegetation}.

Despite attempts to comprehend the dynamics of post-fire recovery for various burn severity levels through remote sensing data while considering topographic and climatic conditions, certain gaps remain in the existing literature. Most studies have concentrated solely on either topographic \citep{ireland2015exploring, leon2012using, Casady2010} or climatic conditions \citep{malak2006fire, hao2022long}. While some studies have examined both factors \citep{viana2017assessment, meng2015effects, chen2022remote, kinoshita2011spatial}, a comprehensive analysis of their individual effects on recovery and the interplay among different factors have not been addressed. Furthermore, prior research has primarily focused on specific fire incidents, overlooking the recovery dynamics for different land cover types. Given the changing climate, it is critical to understand how the climatic condition affects varying landscapes. 

This research aims to understand how the post-fire recovery is jointly influenced by topographic features, post-fire climate conditions, burn severity, and the number of years after the fire. We develop a machine learning model based on the random forest (RF) technique to relate ecosystem recovery with topographic, climatic and fire factors in the Columbia River Basin of the Pacific Northwest (PNW). The specific objectives of this study are: 1) to investigate how topographic variables (elevation, slope, and aspect) and climatic factors (precipitation and temperature) interact to influence vegetation recovery in different burn severity patches over time and 2) to assess whether the interconnected roles of these factors differ across various land cover types. Understanding how different combined factors impact recovery dynamics in different land cover types will provide insights into the varying resilience and response of vegetation to post-fire conditions.   

\section{Materials and Methods}

\subsection{Study area}

\begin{figure*}[htbp]
    \centering
    \begin{subfigure}[t]{1\textwidth}
        \centering
        \caption{}
        \includegraphics[height=3.0in]{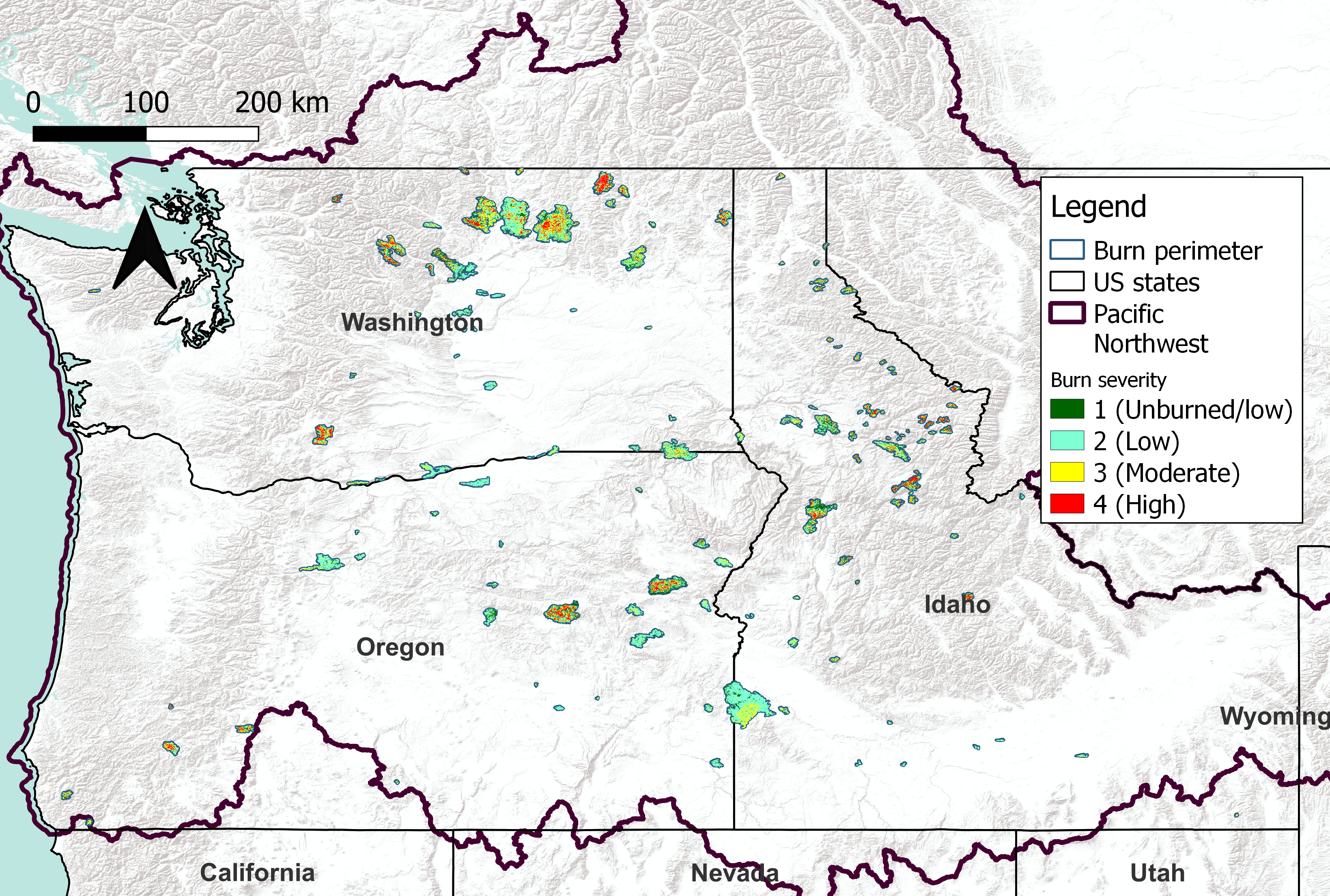}
    \end{subfigure}%
    \\
    \begin{subfigure}[t]{0.4\textwidth}
        \centering
        \caption{}
        \includegraphics[height=2.7in]{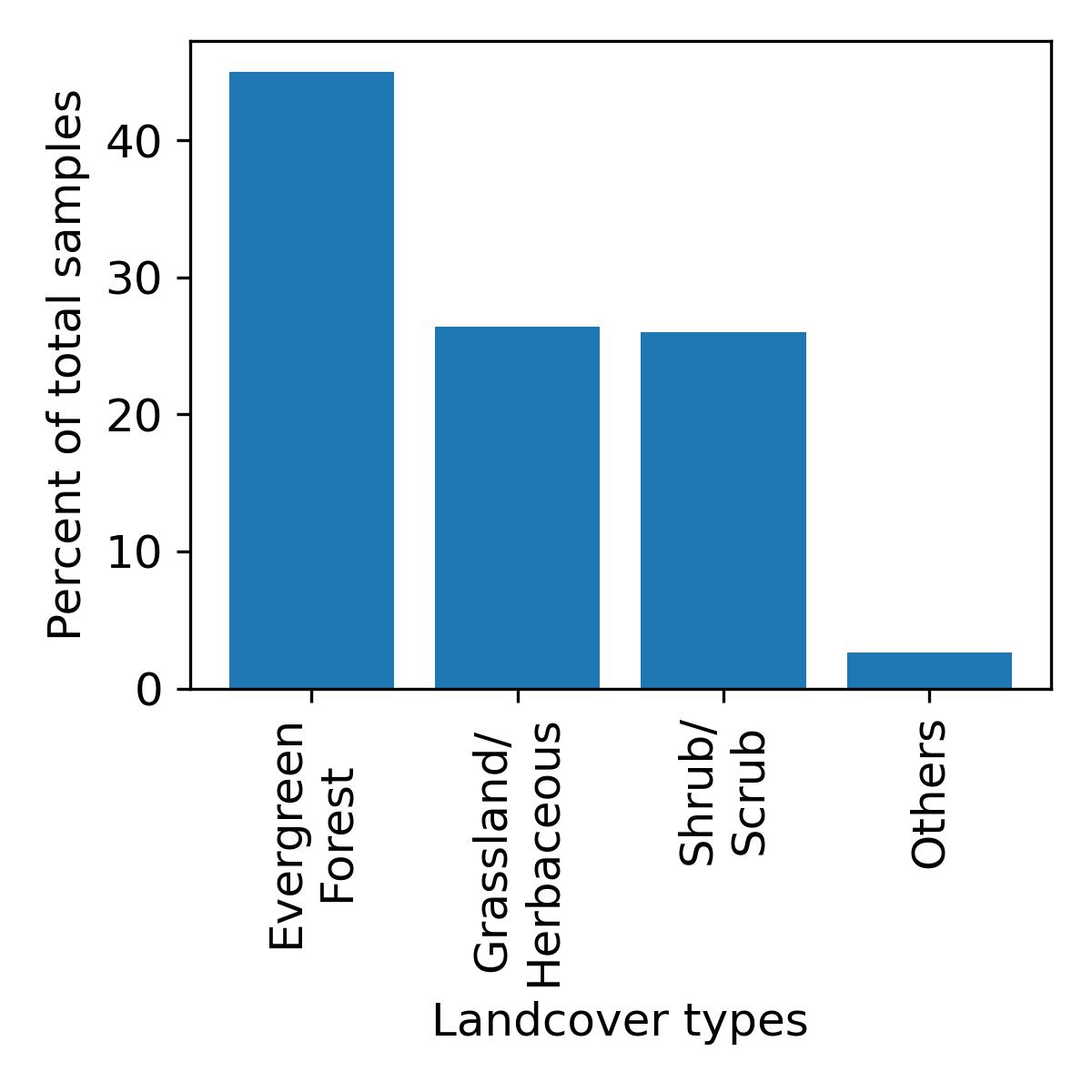}
    \end{subfigure}%
    ~
    \begin{subfigure}[t]{0.5\textwidth}
            \centering
            \caption{}
            \includegraphics[height=2.76in]{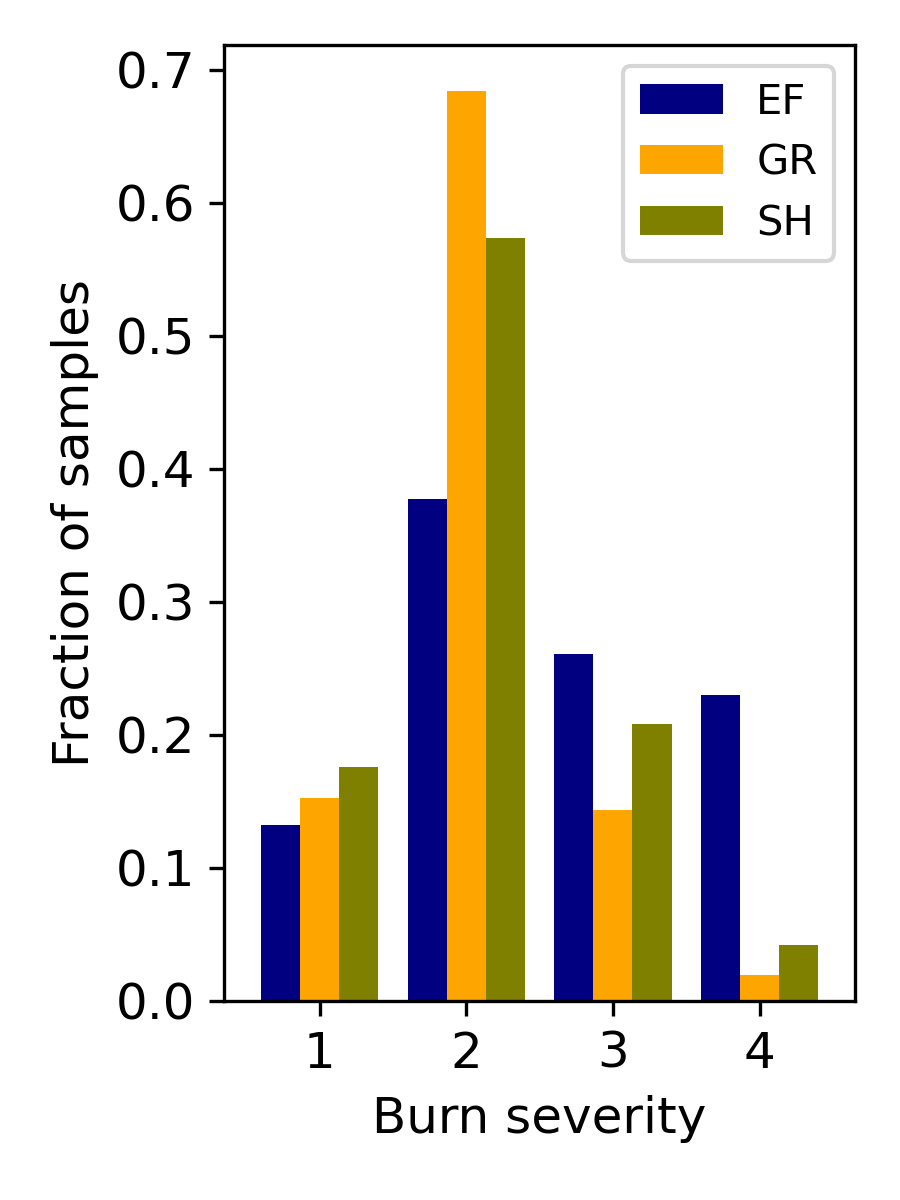}
    \end{subfigure}
    \caption{(a) Burn areas in Washington, Oregon, and Idaho during the 2015 wildfire season, (b) land cover types in the burn areas from NLCD, and (c) the fraction of total gridcells associated with different burn severities from MTBS in dominant land cover types.}
    \label{fig:PNW_burnarea}
\end{figure*}

\begin{figure}
    \centering
    \includegraphics[width=.7\textwidth]{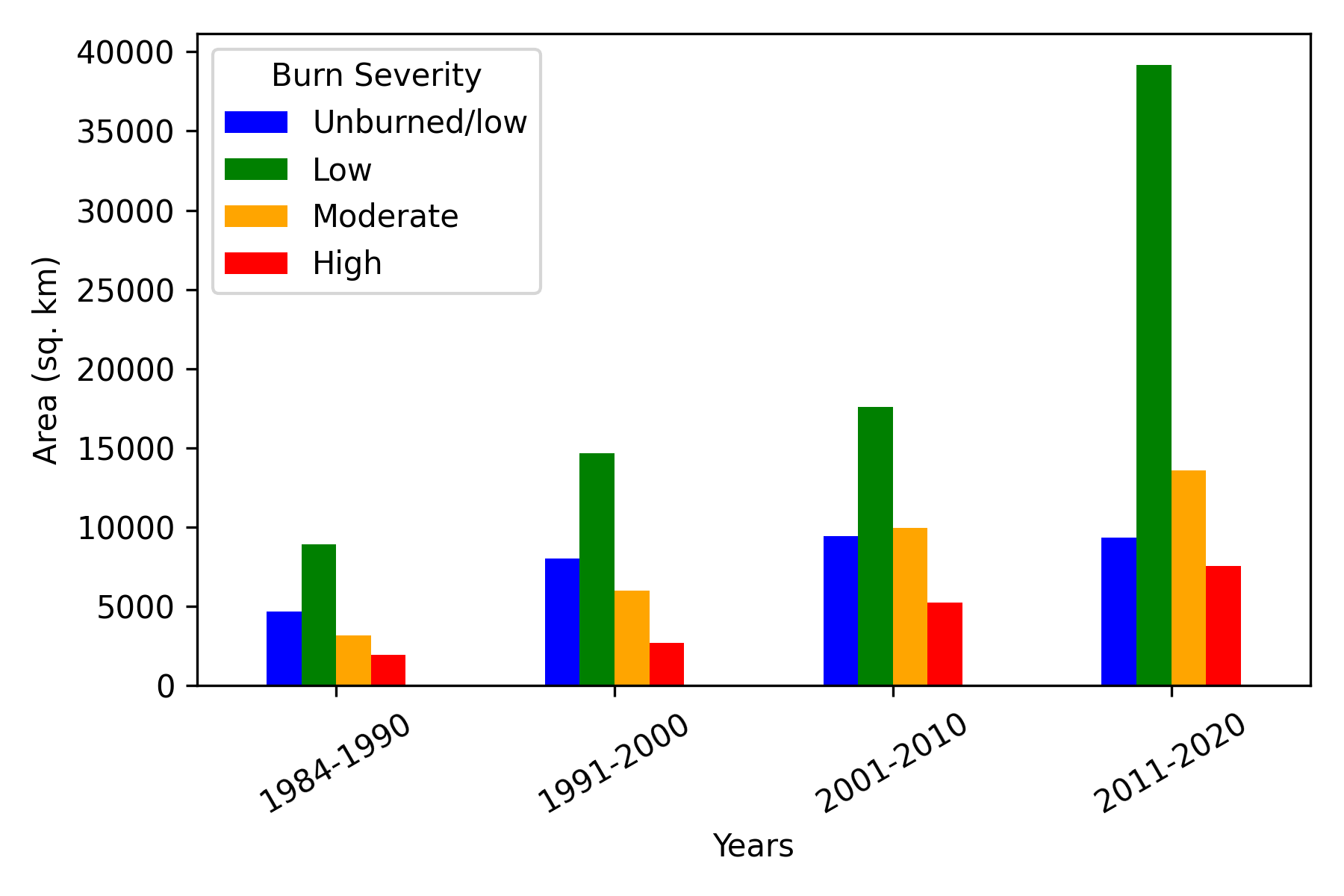}
    \caption{Historical burn area in the PNW region between 1984-2020.}
    \label{fig:historic_burn}
\end{figure}

The study area for this research spans burned areas in Washington, Oregon, and Idaho (Figure~\ref{fig:PNW_burnarea}a) within the Pacific Northwest region. The average annual precipitation in the PNW region is around 900 mm (ranging 171 to 5595 mm spatially), with higher precipitation occurring in the western areas and a drier climate on the east. The average daily maximum temperature ranges spatially from -5.55 to 21.0\textdegree C with a mean of 13.16\textdegree C. Fire occurrence and high burn severity area have escalated in the PNW region over the past decade (Figure~\ref{fig:historic_burn}). During 2011-2020 the total burn area in the study area doubled compared to 1991-2000, while the coverage of high burn severity areas almost tripled. Decreased summer precipitation and rising temperatures due to climate change would continue to escalate the fire occurrences in the PNW \citep{halofsky2020changing}. 

In 2015, the PNW region witnessed higher than average temperatures, with a maximum daily temperature averaged annually over the study area of 14.6\textdegree C. This marked 2015 as the hottest year within the last four decades. The elevated temperatures experienced between 2012 and 2015 led to reduced snow accumulation in the region \citep{mote2016perspectives}. Consequently, widespread drought conditions and increased instances of wildfires occurred in 2015 \citep{marlier20172015}. The wildfires extensively affected the study domain, resulting in the burning of over ten thousand square kilometers of land (Figure~\ref{fig:PNW_burnarea}a). This study aims to better understand the ecological response and recovery of vegetation in the aftermath of the 2015 wildfires. 

\subsection{Input data sources and preprocessing}

\begin{figure}[htbp]
    \centering
    \includegraphics[width=1\textwidth]{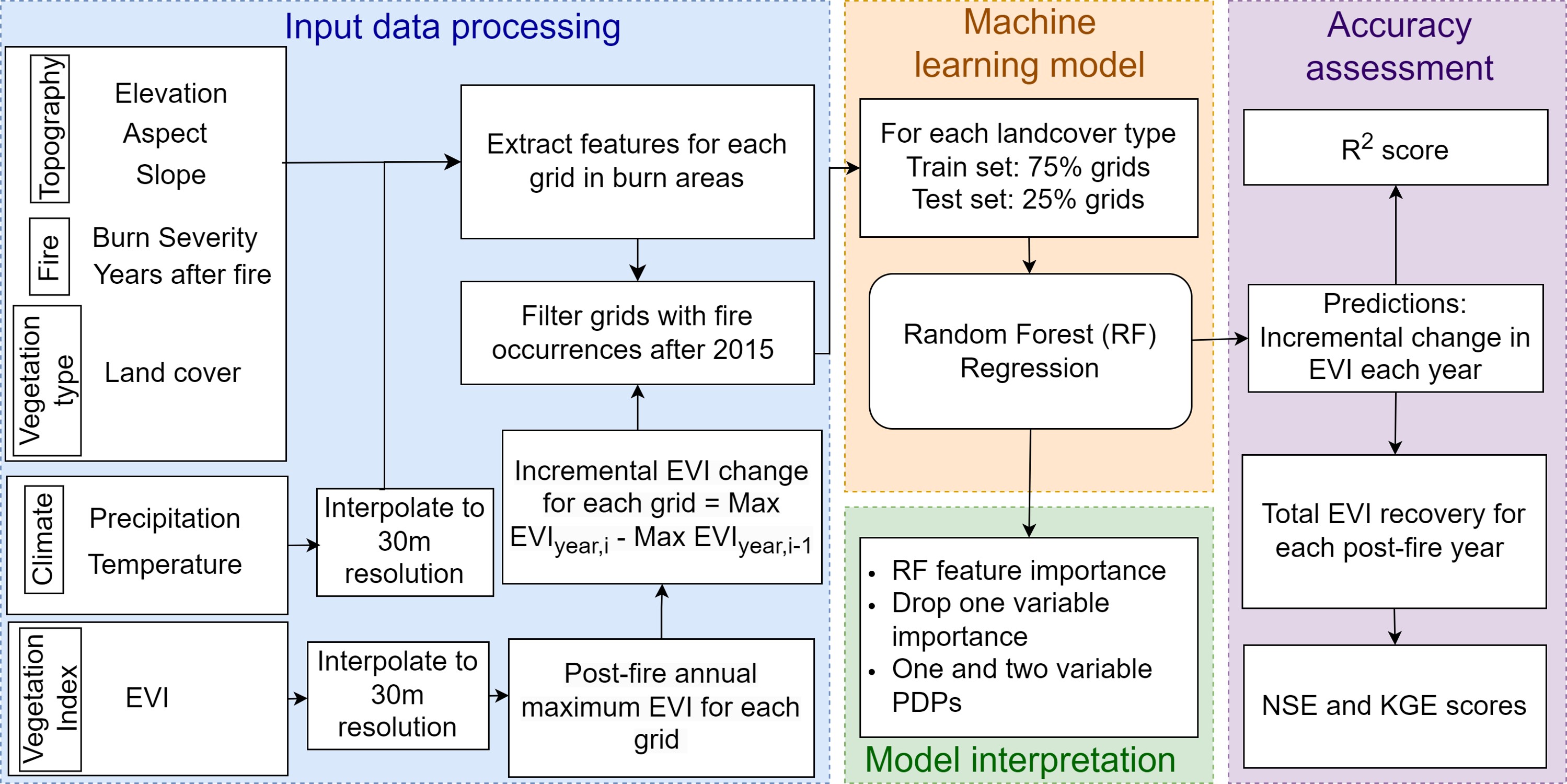}
    \caption{Workflow showing steps for data preprocessing for each of the six post-fire years (2017-2022), RF model training, accuracy assessment, and model interpretation. }
    \label{fig:workflow}
\end{figure}

To understand the spatiotemporal dynamics of vegetation recovery across the study domain, several datasets on topography, climate, land cover, fire characteristics, and vegetation index were gathered and prepared to train the RF model. The complete workflow is shown in Figure~\ref{fig:workflow}.
\subsubsection{Topography}
The three topographic characteristics used in this study (i.e., elevation, slope, and aspect) were derived using the 30 m digital elevation model (DEM), acquired from the U.S. Geological Survey (USGS) \citep{USGS}. From the DEM, slope and aspect rasters were generated for the study region using Quantum Geographic Information System (QGIS) tools \citep{QGIS_software} to account for the steepness and orientation of the terrain, respectively. Aspect data was categorized as north (315\textdegree to 45\textdegree), east (45\textdegree to 135\textdegree), south (135\textdegree to 225\textdegree) and west (225\textdegree to 315\textdegree). 

\subsubsection{Land cover}
To distinguish fire impact on different land cover types, pre-fire land cover data for 2013 was collected at 30 m resolution from USGS National Land Cover Database (NLCD) \citep{NLCD2019}. The percentage of different land cover types in the 2015 burn areas showed the dominant land cover types to be evergreen forests (EF), shrub/scrub (SH), and grassland/herbaceous (GR) (Figure~\ref{fig:PNW_burnarea}(b)). The land cover data was aligned with the DEM grids using the nearest neighbor technique. This process ensured that the land cover and DEM grids were spatially matched for accurate integration and analysis. 

\subsubsection{Climate}

\begin{figure}[htbp]
    \centering
    \includegraphics[width=.6\textwidth]{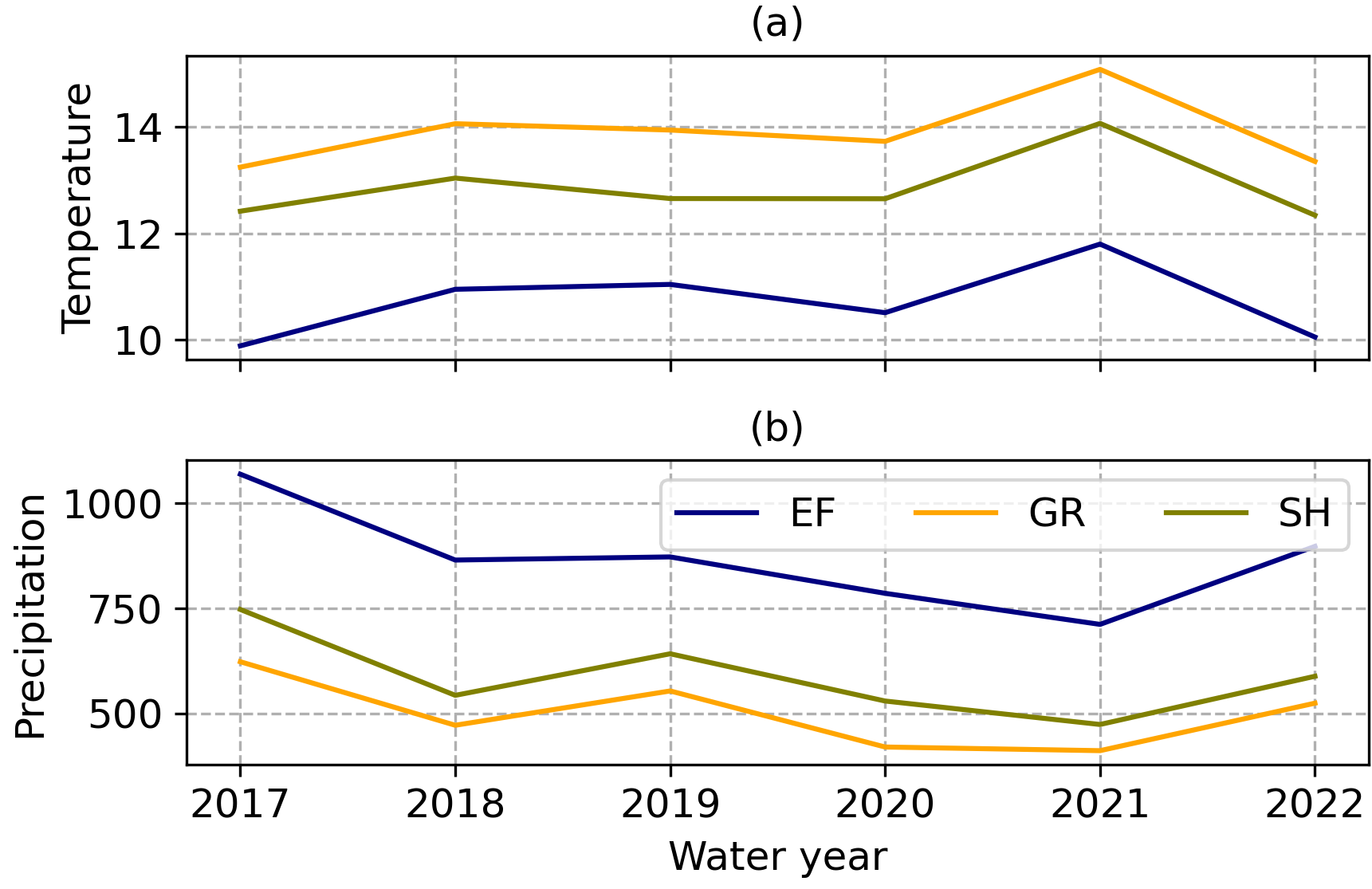}
    \caption{Total precipitation (mm) and mean annual daily maximum temperature (\textdegree C) during the post-fire water years averaged over the burn areas  for evergreen forests (EF), shrub/scrub (SH), and grassland/herbaceous (GR).}
    \label{fig:Post-fire cilmate}
\end{figure}

The monthly total precipitation and daily maximum temperature Daymet data at 1km  were obtained for post-fire years (2016-2022) \citep{https://doi.org/10.3334/ornldaac/2131}. Precipitation and temperature data were interpolated to 30 m to match the grids in DEM. Figure~\ref{fig:Post-fire cilmate} shows the annual total precipitation and average daily maximum temperature for the three land cover types in the burn areas during post-fire water years (a 12-month period starting from October 1 in a year to September 30 in the following year).

\subsubsection{Burn severity}
Geospatial datasets on the burn area perimeter and 30 m resolution burn severity patches for the study region were obtained from the Monitoring Trends in Burn Severity (MTBS) \citep{MTBS2022} database. MTBS uses Landsat data to calculate the pre-fire and post-fire normalized burn ratio (NBR). The difference between the pre- and post-fire NBR (dNBR) is used to delineate burn areas and partitioned to get five burn severity classes: (1) unburned to low, (2) low, (3) moderate, (4) high, and (5) increased greenness \citep{eidenshink2007project}. Burn severity for the 2015 fires in the study domain is shown in Figure~\ref{fig:PNW_burnarea}(a). Similar to land cover data, burn severity grids were also aligned to match the DEM.

Burn severity levels for the three land cover types are shown in (Figure~\ref{fig:PNW_burnarea}(c)). The majority of GR and SH grids, 68\% and 57\%, respectively, experienced low burn severity, while only 2\% and 4\% of the grids, respectively, burned with high severity. Compared to GR and SH, burns in EF grids had larger percentage (23\%) with high severity and a lower percentage (38\%) with low severity. 

\subsubsection{Vegetation index: EVI}
For the purpose of tracking the post-fire vegetation regrowth, EVI data (MOD13Q1.006) was obtained from the Terra Moderate Resolution Imaging Spectroradiometer (MODIS) \citep{MOD13Q12021}. The MODIS EVI product has a spatial resolution of 250 m and is calculated at a temporal scale of 16 days as \citep{liu1995feedback}
\begin{equation}\label{eq_EVI}
    EVI=G\times\frac{NIR-RED}{NIR+C_{1}\times RED-C_{2}\times BLUE+L}\
\end{equation}
where NIR, RED, and BLUE are near-infrared, red, and blue band reflectances, respectively. $G$ and $L$ are gain factor and canopy background adjustment factor, respectively. $C_{1}$ and $C_{2}$ represent the coefficients of the aerosol resistance term, which employs the blue band to correct for aerosol impacts on the red band. The values for $G$, $L$, $C_{1}$, and $C_{2}$ used in the MODIS EVI calculation are 2.5, 1, 6, and 7.5, respectively. The 250 m EVI data were interpolated to 30 m to ensure consistency in resolution and grid coordinates with DEM. 

\subsubsection{Input dataset generation}
For each 30 m grid in the burn area, the corresponding elevation, slope, aspect, precipitation, temperature, burn severity, land cover, and EVI values were extracted. The total precipitation and median of the daily maximum temperature were computed for each grid during a water year. Additionally, the maximum EVI values were determined for each post-fire water year between 2016 and 2022 for each grid cell. The incremental EVI change for a specific post-fire year was then calculated as the difference between the maximum EVI value for the year of interest and the maximum EVI value from the previous year. The elapsed time since fire occurrence was determined by calculating the difference in years between the year of interest and the year the fire took place (2015). The values for years after the fire range from 2 to 7. The incremental EVI change for the year 2016 was omitted from the analysis due to its immediate occurrence following the fire, which resulted in a decrease in EVI values rather than recovery. Thus, input datasets were generated for the three dominant land cover types (EF, SH, and GR) containing input features such as elevation, slope, aspect, annual precipitation and temperature, burn severity, and years after fire with incremental change in EVI as the target variable (Table~\ref{table:Features}). Subsequent fire occurrences in the 2015 burn areas would cause vegetation mortality, reducing EVI values. Therefore, grid cells experiencing fire after 2015 were excluded from the analysis to focus on areas that did not experience additional fire incidents.

\begin{table}[htbp]
\centering
\caption{Input and target features to the RF model.}
\begin{tabular}{ |c|c|c|c|}
 \hline
 \multicolumn{2}{|c|}{Variables} & Units & Variability\\ 
 \hline\hline
 \multirow{7}{3em}{Inputs} & Elevation & m & Spatial \\ 
 & Slope & degree & Spatial \\ 
 & Aspect & degree & Spatial \\
 & Precipitation & mm & Spatial and temporal \\
 & Temperature & \textdegree C & Spatial and temporal \\
 & Burn severity & n/a & Spatial \\
 & Years after fire & years & Temporal \\
 \hline\hline
 Target & Incremental change in EVI & n/a & Spatial and temporal \\
 \hline
\end{tabular}
\label{table:Features}
\end{table}

\subsection{Random Forest regression}
RF regression via the sklearn python package \citep{scikitlearn,RanForReg2023} was used to explore the non-linear relationship between the input and target variables. The RF algorithm constructs an ensemble of decision trees. Each tree is built using a random subset of the training data. For splitting, a random subset of features is considered at each node of the tree \citep{breiman2001random}. This randomness in data and feature selection reduces overfitting. Each tree in the ensemble makes predictions independently, which are averaged to calculate the final prediction. RF hyperparameters, such as the number of trees, maximum depth of trees, number of features considered at splitting, minimum samples per leaf, and minimum samples for splitting, can be adjusted to optimize model performance and improve predictive accuracy.
 
\subsection{Model training}
The RF model was trained to learn the relationship between the input features regarding topographic, climate, and fire-related characteristics and the desired output, incremental change in EVI on an annual basis. From datasets on each land cover type, 75\% of the grids were used for training the model and the other 25\% for testing. The main hyperparameters in the RF model are (1) number of trees (n\_estimators), (2) number of features considered at splitting (max\_features), and (3) maximum depth (max\_depth). These were tuned using GridSearchCV with a 5-fold cross validation in the python sklearn package \citep{GridSearchCV_2023} on the training set. GridSearchCV uses k-fold cross validation on a dataset and iterates over provided possible combinations of parameters to find the best set of hyperparameters. Although increasing the number of trees and maximum depth in the model can lead to improved scores, it should be noted that there is a trade-off. Beyond a certain point, further increasing these parameters may result in longer training times without a significant improvement in model performance. In the final model training, optimum hyperparatemeters were selected based on model score improvement and computational time. Values for n\_estimators, max\_depth, and max\_features were set to 50, 50, and 6, respectively.

\subsection{Accuracy assessment of the model}
The model performance on train and test sets was evaluated using $R^2$ scores between the observed $y$ and predicted $\hat{y}$ values of incremental EVI changes as 
\begin{equation}
    R^2 = 1-\frac{\sum(y-\hat{y})^2}{\sum(y-\bar{y})^2}
\end{equation}
where $\bar{y}$ is the mean of the observed values for incremental EVI changes. $R^2$ values can range from $-\infty$ to 1, where higher scores indicate good agreement between observations and predictions. Additionally, the total recovery for each grid was calculated for each post-fire year, from 2017 to 2022. This calculation involved aggregating the incremental EVI changes starting in 2017 up to the specific year of interest. The total recovery calculated from observations and predictions was compared by estimating Nash–Sutcliffe and Kling-Gupta efficiency (NSE and KGE) scores to determine whether the model-predicted incremental recoveries add up to the observed total recovery. NSE and KGE were calculated for each post-fire year over all burn grids to assess the RF model's ability to capture the spatial pattern of recovery. Higher scores (close to 1) suggest a better match, while lower scores indicate a discrepancy between the observed and predicted values. NSE and KGE were calculated over $N$ number of grid cells for each year, $i$, as
\begin{equation}
    NSE_i = 1-\frac{\sum_{n=1}^{n=N}(y_{i,n}-\hat{y}_{i,n})^2}{\sum_{n=1}^{n=N}(y_{i,n}-\bar{y}_{i})^2}\
\end{equation}
where $y_i$, $\hat{y}_{i}$, and $\bar{y}_{i}$ are the observed, predicted, and average of the observed total EVI recovery, respectively, and
\begin{equation}
    KGE_i = 1-\sqrt{(r_i-1)^2+(\frac{\sigma(\hat{y}_{i})}{\sigma(y_{i})}-1)^2+(\frac{\mu(\hat{y}_{i})}{\mu({y}_{i})}-1)^2}\
\end{equation}
where $r_i$ is Pearson coefficient between observations, $y_i$, and predictions, $\hat{y}_{i}$. $\sigma$ and $\mu$ are the standard deviation and average of the total EVI recovery.

\subsection{Model interpretability}

\subsubsection{RF feature importance}
Besides predicting a target variable, RF provides the relative importance of features, called the Gini importance. In sklearn, the feature importance is calculated as a normalized total reduction in mean squared errors by each feature while constructing trees in the forest. A higher value of importance score indicates a greater importance for the specific feature. The RF feature importance is measured during tree construction. However, the individual impact of features on predicting the target variable for the train and test sets is not directly addressed. Moreover, the RF generated relative feature importance can exhibit bias towards variables with high cardinality.

\subsubsection{Drop one variable importance}
To gather a quantitative understanding of the importance of each feature in model predictions, the drop one variable approach was used. The RF model was trained by dropping each of the seven input features one at a time. $R^2$ scores were calculated on both the training and testing datasets. Scores from the drop one variable approach were compared with the baseline scenario, where RF was trained with all features, to assess the impact of excluding each feature on the RF model's performance.

\subsubsection{One and two variable partial dependence plots (PDP)}
While the feature importance matrices reveal how important a feature is in improving model performance, they do not display the relationship between target and individual feature. Therefore, PDPs were used to help understand black-box model predictions by graphically showing the relationship between one or two specific input features and the target variable \citep{friedman2001greedy}. One variable PDPs illustrate how the predicted output changes as a particular input feature varies, while holding all other features constant at their average values. Two-variable PDPs help with visualizing the interaction and joint influence of two features on the model predictions while marginalizing the impact of all other input features. PDPs for each input feature and pairs of features were generated using sklearn's partial\_dependence method \citep{PartialDependence_2023} on the training set. 

\section{Results and Discussion}
\subsection{Pre and post-fire EVI values}
\begin{figure}[htbp]
    \centering
    \includegraphics[width=.65\textwidth]{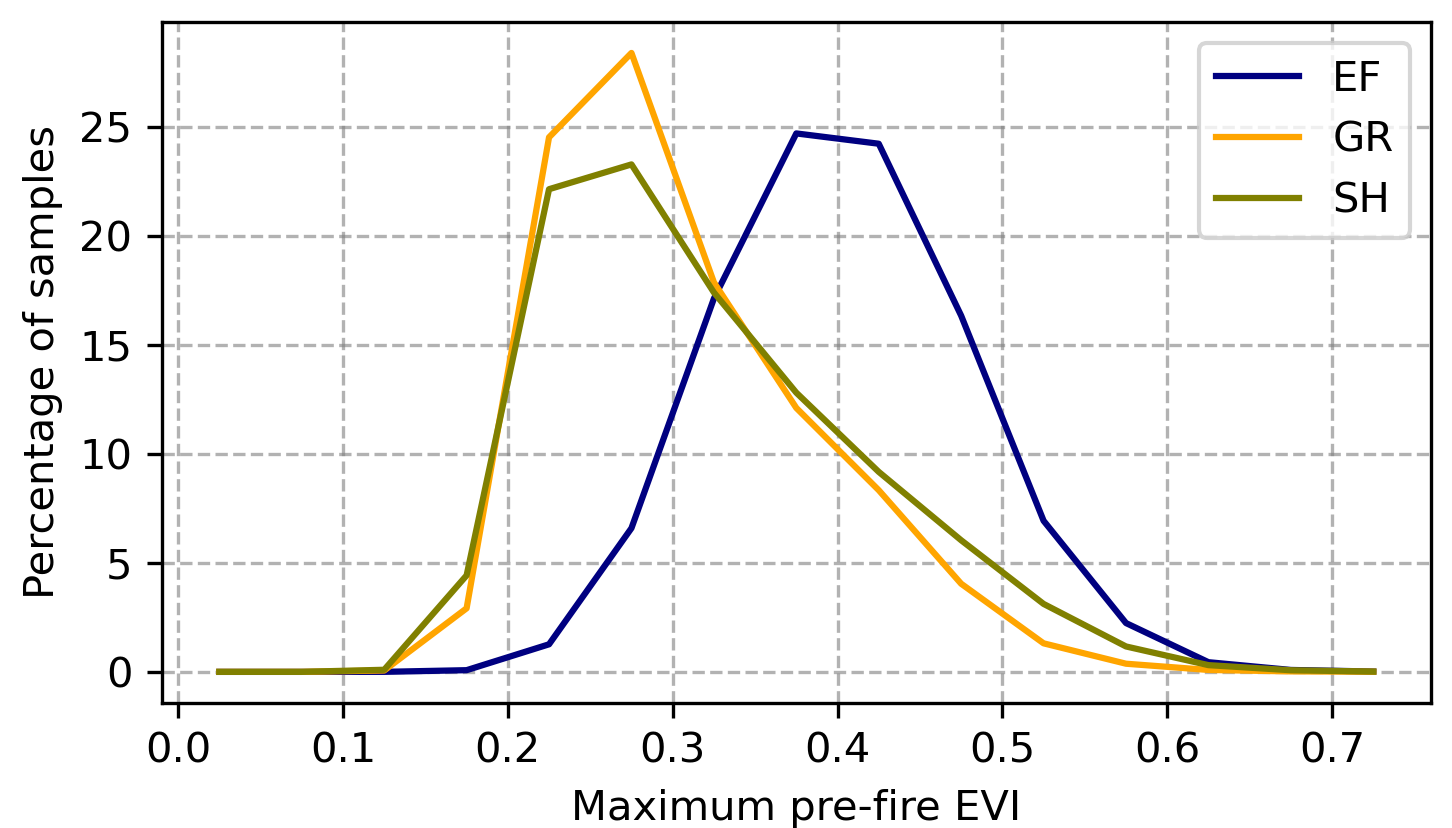}
    \caption{Average annual maximum EVI during five pre-fire years for EF, GR, and SH land cover types.}
    \label{fig:Pre-fire EVI}
\end{figure}

\begin{figure}[htbp]
    \centering
    \includegraphics[width=1\textwidth]{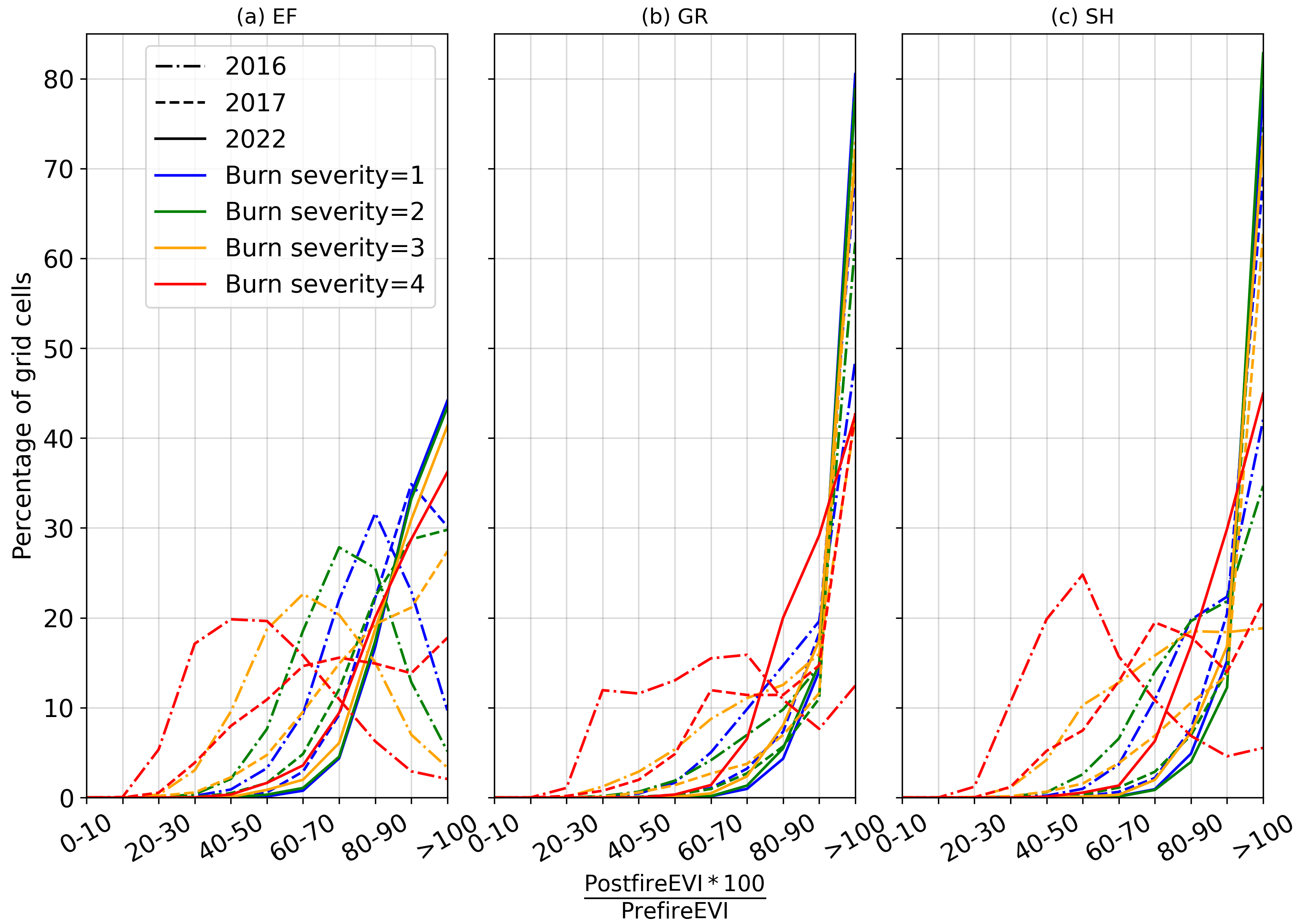}
    \caption{Post-fire decline and gradual recovery of annual maximum EVI values for (a) evergreen forests (EF), (b) grassland/herbaceous (GR), and (c) shrub/scrub (SH).}
    \label{fig:Post-fire EVI}
\end{figure}

Figure~\ref{fig:Pre-fire EVI} describes the maximum EVI values for the burned grids averaged over five pre-fire years (2010-2014) for the three land cover types. The annual maximum EVI for 82\% of the EF grids was within 0.35-0.50, while around 80\% of SH and GR grids had maximum EVI values within 0.25-0.40 (Figure~\ref{fig:Pre-fire EVI}). Post-fire annual maximum EVI values for the first (2016), second (2017), and seventh (2022) years after the fire are shown in Figure~\ref{fig:Post-fire EVI} as percentages of the pre-fire maximum EVI on corresponding grid cells. During the first year (2016, represented with dash-dot lines), immediately following the 2015 fires, the EVI values dropped for all the land cover types. Lower burn severity (blue) resulted smaller decrease in EVI values, while high burn severity (red) was associated with a larger decrease. The decrease in EVI values was larger for EF land cover than for SH and GR. 

In 2016, post-fire EVI values for burn severity level 1 dropped between 60-90\% of the pre-fire value in 63\% of EF grids and nearly 30\% of SH and GR grids. 42 and 47\% of SH and GR grids, respectively, maintained EVI values similar to pre-fire levels. For burn severity 4, 83\%, 82\%, and 68\% of EF, SH, and GR grids, respectively, experienced EVI values that decreased to 30-80\% of pre-fire values. During the second year after fire (2017, represented with dashed lines), the first year for vegetation regeneration after the fire disturbances, 65\% of EF and roughly 88\% of SH and GR grids with burn severity 1 reached 90\% or more of pre-fire EVI values. In contrast, 32, 36, and 57\% of EF, SH, and GR grids, respectively, with burn severity 4 reached >90\% of pre-fire EVI. In 2022 (solid lines), 94\% of the SH and GR grids recovered to >90\% of pre-fire values, while 78\% of EF grids experienced a similar level of recovery for burn severity 1. The number of recovered grids was smaller for high burn severity grids, with 65, 75, and 72\% of EF, SH, and GR grids, respectively, recovered to >90\% of pre-fire EVI. 

As high burn severity grids undergo larger decreases in EVI, they take longer to recover and their recovery dynamics depend on the post-fire climate and landscape characteristics. This study investigates the annual progression of EVI recovery in connection with these influencing factors.

\subsection{Predicting EVI incremental change and total recovery }
\begin{figure}[htbp]
    \centering
    \includegraphics[width=1\textwidth]{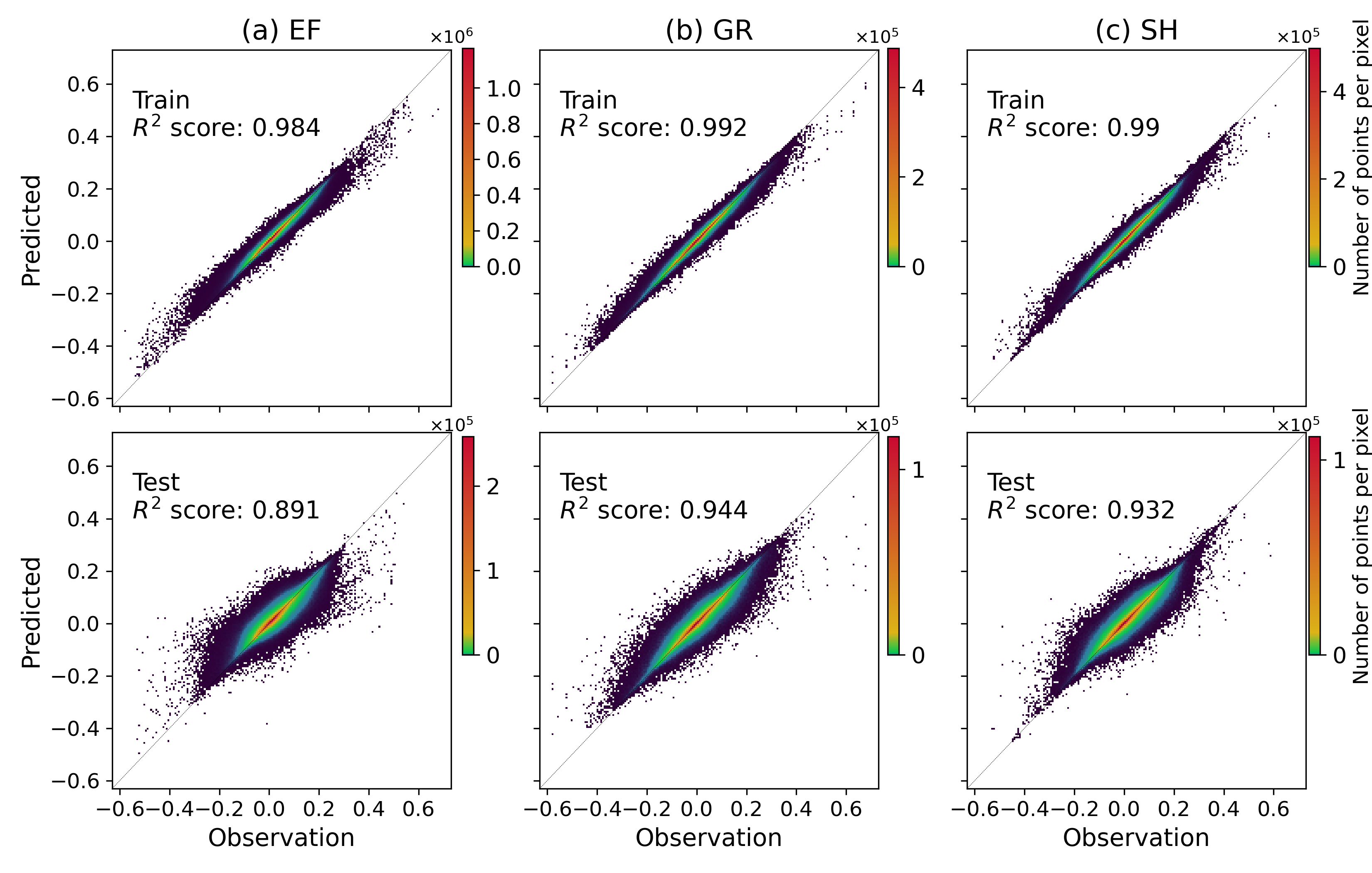}
    \caption{Observed and predicted incremental changes in EVI for different land cover types (a) EF, (b) GR, and (c) SH. }
    \label{fig:obs_vs_pred}
\end{figure}

The RF model was able to accurately predict the observed incremental changes in EVI for all three land cover types for both training and test sets (Figure~\ref{fig:obs_vs_pred}). For the train set, the $R^2$ scores were approximately 0.99 for all three land cover types. These high $R^2$ scores indicate a strong linear relationship between the observed and predicted incremental changes in EVI within the training data. 

When evaluating the model's performance on the test set, the $R^2$ scores showed a slight decrease compared to the train set. However, they still remained quite high, ranging from 0.891 to 0.944 across the three land cover types. These scores indicate that the RF model effectively predicted incremental changes in EVI for unseen data without overfitting. 

\begin{figure}[htbp]
    \centering
    \includegraphics[width=.7\textwidth]{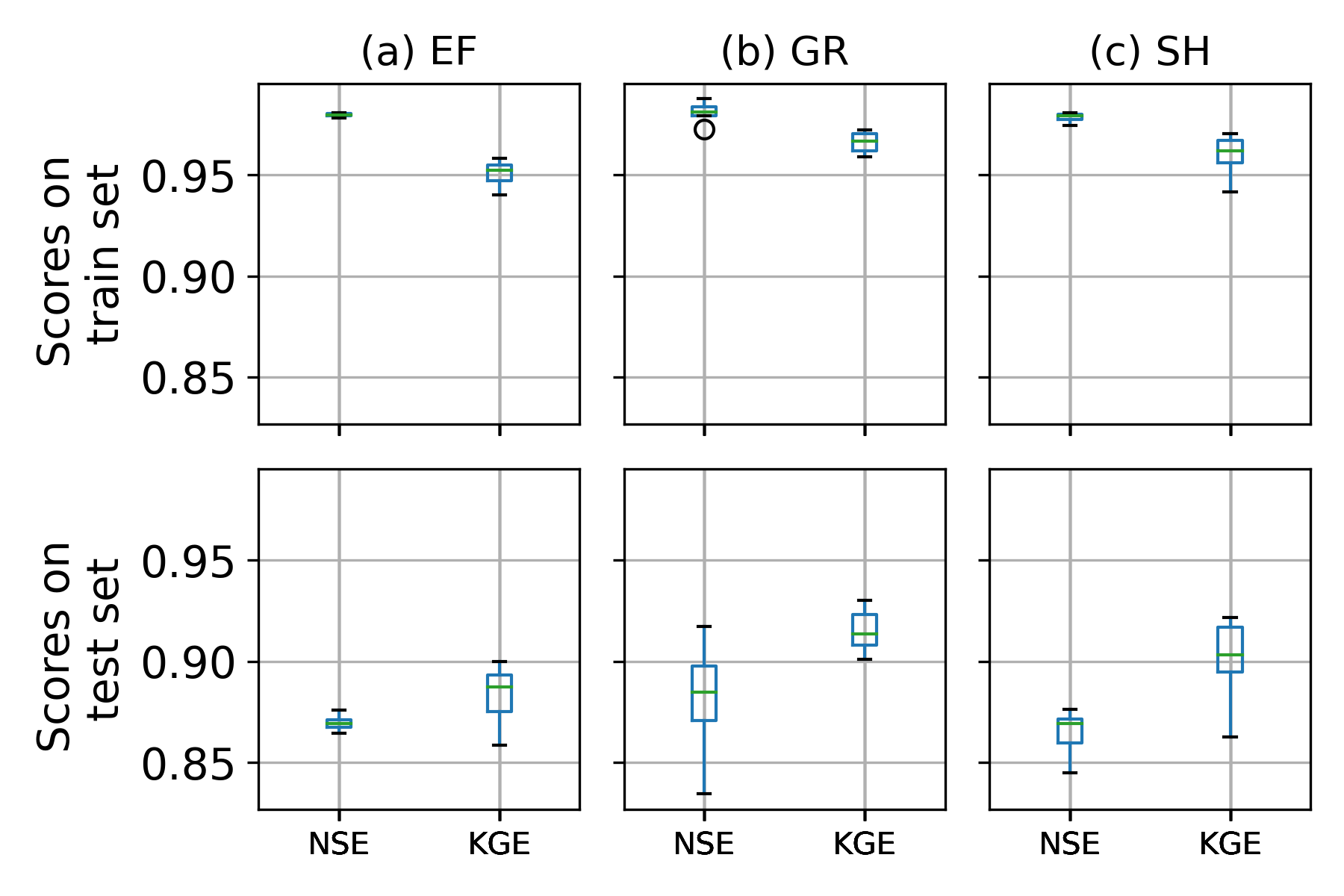}
    \caption{NSE and KGE values for the train and test sets for the six post-fire years and different land cover types (a) EF, (b) GR, and (c) SH. }
    \label{fig:NSE_KGE}
\end{figure}

In the training set, the NSE and KGE values were consistently above 0.98 and 0.94, respectively, indicating excellent agreement between the observed and predicted values across land cover types and post-fire years (Figure~\ref{fig:NSE_KGE}). In the testing set, the average NSE values over the six-year period for all land cover types were around 0.87, while the average KGE values were between 0.88 to 0.91, further indicating the model's capability to capture spatial variability in the data. The yearly total recovery was not a direct output from the RF model. Instead, it was derived by aggregating the incremental EVI changes over the post-fire years. The relatively high values of NSE and KGE indicated that the RF model captured the patterns and trends in the incremental EVI changes, allowing for an accurate calculation of the yearly total recovery. Therefore, the RF model can be employed to analyze the influence of various features on the model outputs and gain insights into their effects.

\subsection{RF feature importance}
\begin{figure}[htbp]
    \centering
    \includegraphics[width=.7\textwidth]{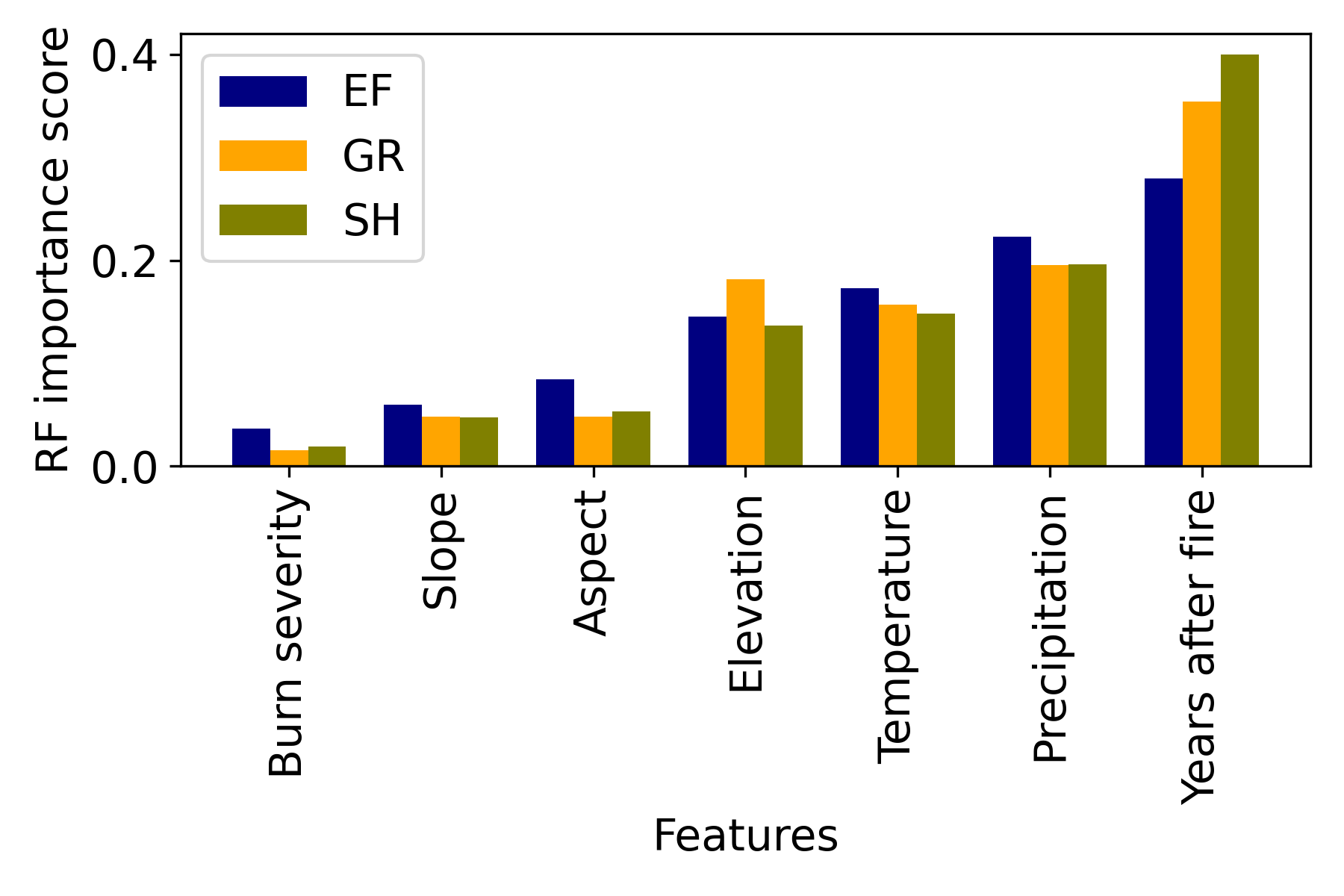}
    \caption{RF feature importance for different land cover types EF, GR, and SH. }
    \label{fig:RF_importance}
\end{figure}

The relative importance metrics for features from RF (Figure ~\ref{fig:RF_importance}) showed that the years after the fire was the most important for all three land cover types, followed by precipitation. For EF and SH, temperature held the position of the third most important feature, while for GR, elevation took the third spot. Although years after fire had the highest importance score for all land cover types, its scores were higher for GR and SH than EF. The rest of the features showed higher scores for EF compared to GR and SH, suggesting that these features have a more substantial influence in predicting incremental EVI changes for EF. The least important feature, burn severity, had the lowest score for GR (0.015), which could be indicative of the lack of variation in burn severity for GR in the studied region. GR and SH grids mostly burned at low severity (57-68\%), with a small fraction burned at high severity ($<$4\%) (Figure~\ref{fig:PNW_burnarea}(c)).

\subsection{Drop one variable importance}
\begin{figure}[htbp]
    \centering
    \includegraphics[width=.7\textwidth]{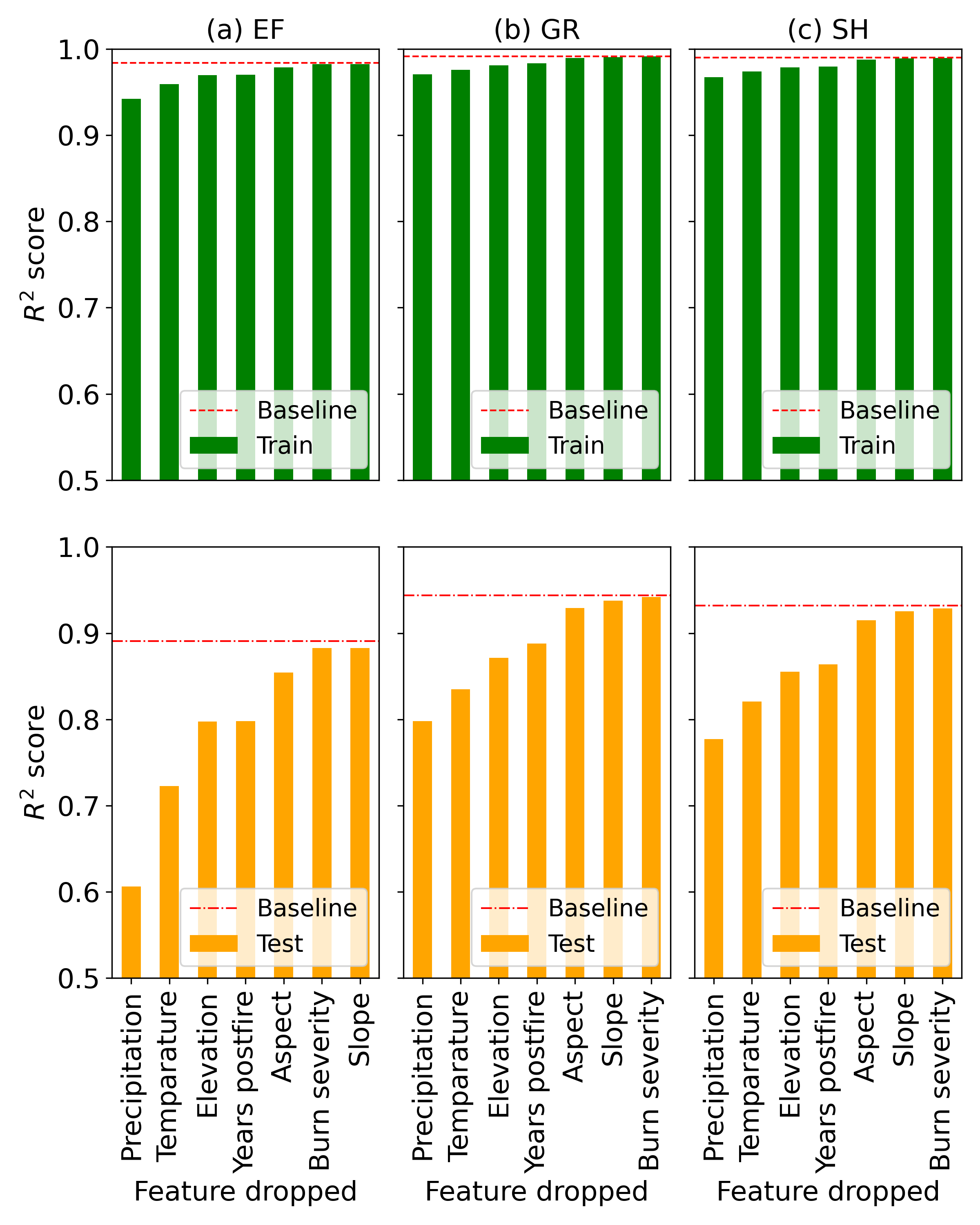}
    \caption{$R^2$ scores for the training (top row) and testing sets (bottom row) for three land cover types with one variable dropped in training. The reference score represents the train and test scores for the model trained with all variables.}
    \label{fig:drop_one_var}
\end{figure}

The drop one variable importance for the seven input features revealed that the model predictions were strongly influenced by climate variables for all land cover types (Figure~\ref{fig:drop_one_var}). Although RF feature importance indicated that years after fire was the most significant feature, the increased emphasis on climate variables in the drop one variable importance underscores their crucial role in determining moisture availability. Excluding precipitation and temperature caused a drop in training $R^2$ scores ranging 0.023-0.042 and 0.016-0.025, respectively, compared to the baseline training $R^2$ scores (Figure~\ref{fig:obs_vs_pred} and ~\ref{fig:drop_one_var}). Although the drops in training scores were relatively small, the impact on the model's performance became more evident when evaluating the test $R^2$ scores. In the case of the test set, excluding precipitation led to a drop in $R^2$ scores between 0.146 and 0.285, while excluding temperature resulted in a decrease between 0.11 and 0.169 for the three land cover types. These larger drops in the test $R^2$ scores indicate that the absence of precipitation and temperature as input features had a more significant effect on the model's performance when using unseen data. 

Among the topographic features, elevation was found to be the most influential, reducing test $R^2$ scores by between 0.073 and 0.093. The changes in the train and test $R^2$ scores caused by slope and burn severity were relatively minimal. Reduction in test scores by excluding slope and burn severity ranged from 0.007-0.008 and 0.002-0.008, respectively.

\subsection{One variable PDP}
\begin{figure}[htbp]
    \centering
    \includegraphics[width=1\textwidth]{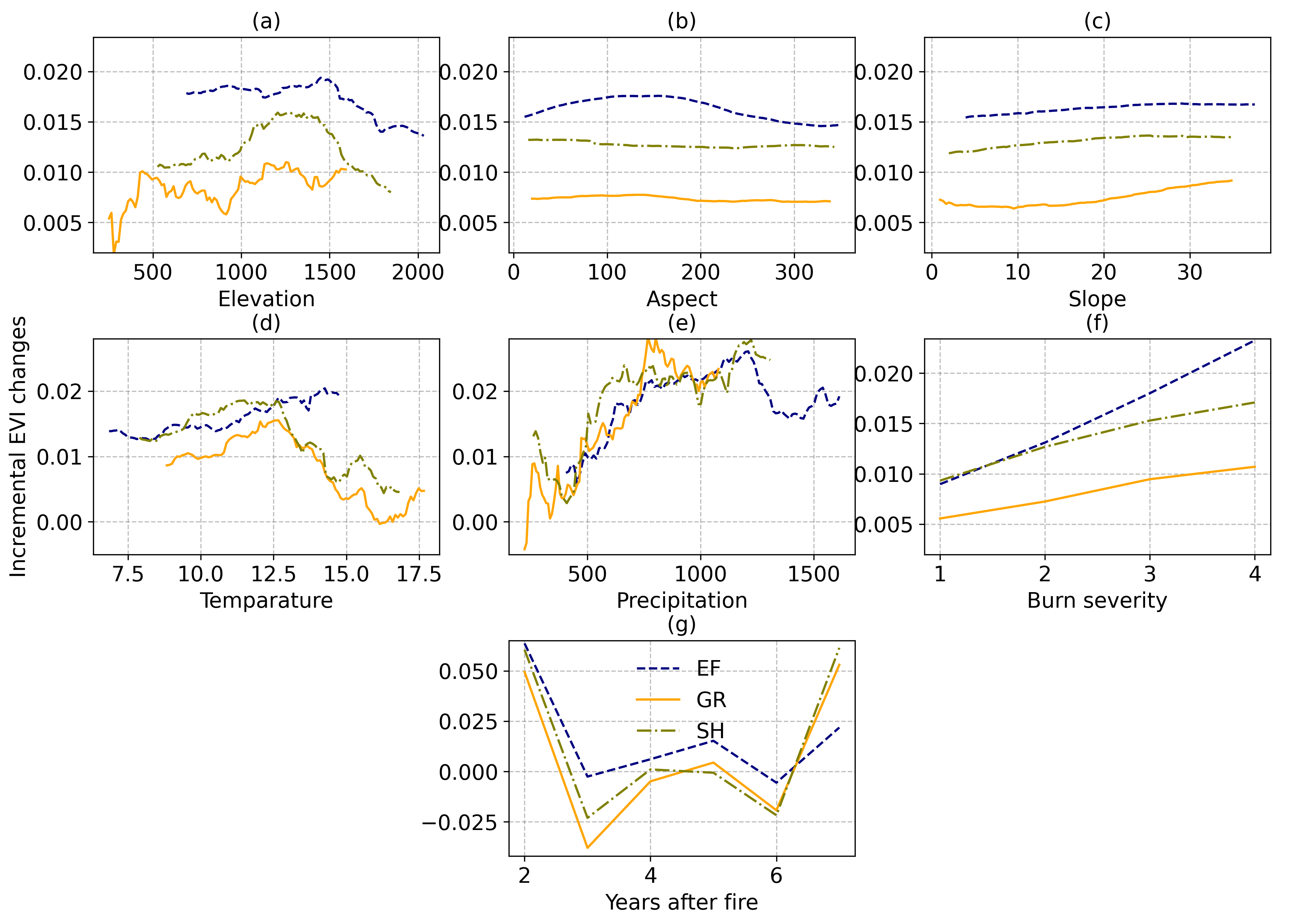}
    \caption{One variable PDPs for three land cover types and variables (a) elevation, (b) aspect, (c) slope, (d) temperature, (e) precipitation, (f) burn severity, and (g) years after fire.}
    \label{fig:PDP_1var}
\end{figure}

The PDPs for topographic variables exhibit varying responses across land cover types(Figure~\ref{fig:PDP_1var}a-c). In the case of EF, the incremental change in EVI remains relatively stable up to an elevation of 1500m and then decreases afterward. For SH, the incremental change in EVI increases with an elevation up to 1000m, remains stable until 1500m, and then decreases after 1500m. This indicates potential elevation ranges where SH zones exhibit higher recovery rates. The PDP values for GR show a fluctuating relationship with elevation, however, an increasing trend is generally observed. Aspect PDPs revealed that variation in aspect values affected incremental EVI changes only for EF and remained comparatively flat for SH and GR. Grids with EF that showed higher incremental changes in EVI appeared mostly on south and east-facing aspects, while lower recovery was predicted on north and west-facing aspects. For SH and GR, the PDP values remain flat with varying aspect. Slope PDPs showed an association of higher slope with higher PDP values for GR, whereas for EF and SH they remain comparatively flat.

Temperature PDPs for EF exhibited increasing PDP values with increasing temperature, whereas they show a gradual increase followed by a decreasing trend after annual median maximum temperatures of 12.5\textdegree C for SH and GR (Figure~\ref{fig:PDP_1var}d). This could potentially be indicative of arid environments due to high temperature hindering vegetation recovery.  For all land cover types, increasing precipitation resulted increasing incremental EVI changes, indicating higher recovery (Figure~\ref{fig:PDP_1var}e). Across all land cover types, a higher incremental change in EVI value was associated with higher burn severity (Figure~\ref{fig:PDP_1var}f). As larger drops in EVI values occurred in high burn severity grids (Figure~\ref{fig:Post-fire EVI}), there was additional room for increases. Low burn severity grids reached pre-fire EVI values faster than high burn severity grids. Incremental EVI changes varied from year to year, potentially depending on the climate conditions of the year (Figure~\ref{fig:PDP_1var}g). For all land cover types, maximum PDP values appeared during the second year after the fire (i.e., 2017), which had the highest total precipitation during post-fire period (Figure ~\ref{fig:Post-fire cilmate}). The sixth post-fire year (2021) had the least rainfall and highest temperature for the burn areas. Therefore, all land covers experienced negative PDP values, indicating increased vegetation mortality compared to the prior year.

\subsection{Two variable PDP}
\begin{figure}[htbp] 
    \centering
    \includegraphics[width=1\textwidth]{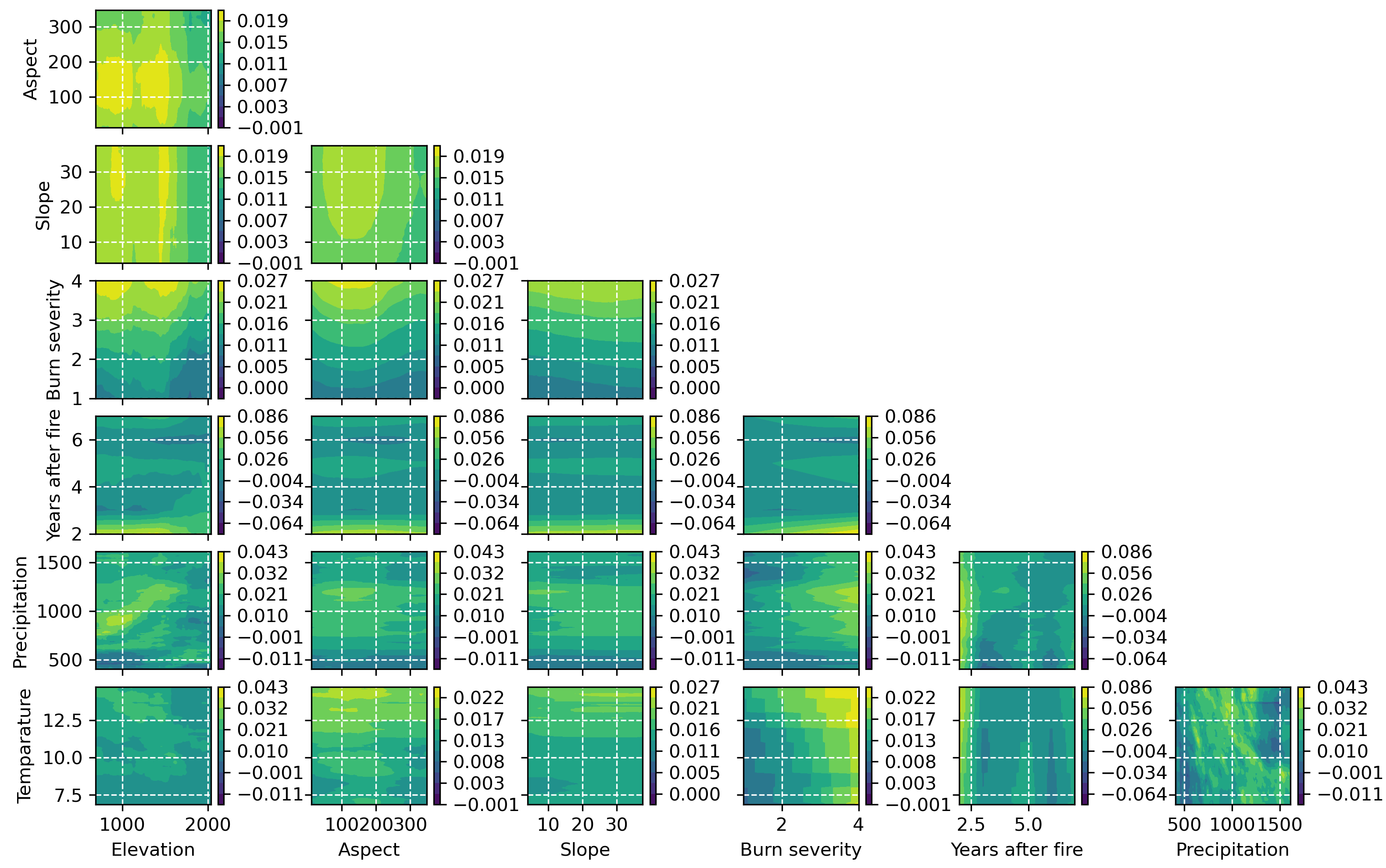}
    \caption{Two variable PDPs for EF showing the joint influence of variable pairs on predicting the target variable, incremental changes in EVI.}
    \label{fig:PDP_2var_EF}
\end{figure}

\begin{figure}[htbp]
    \centering
    \includegraphics[width=1\textwidth]{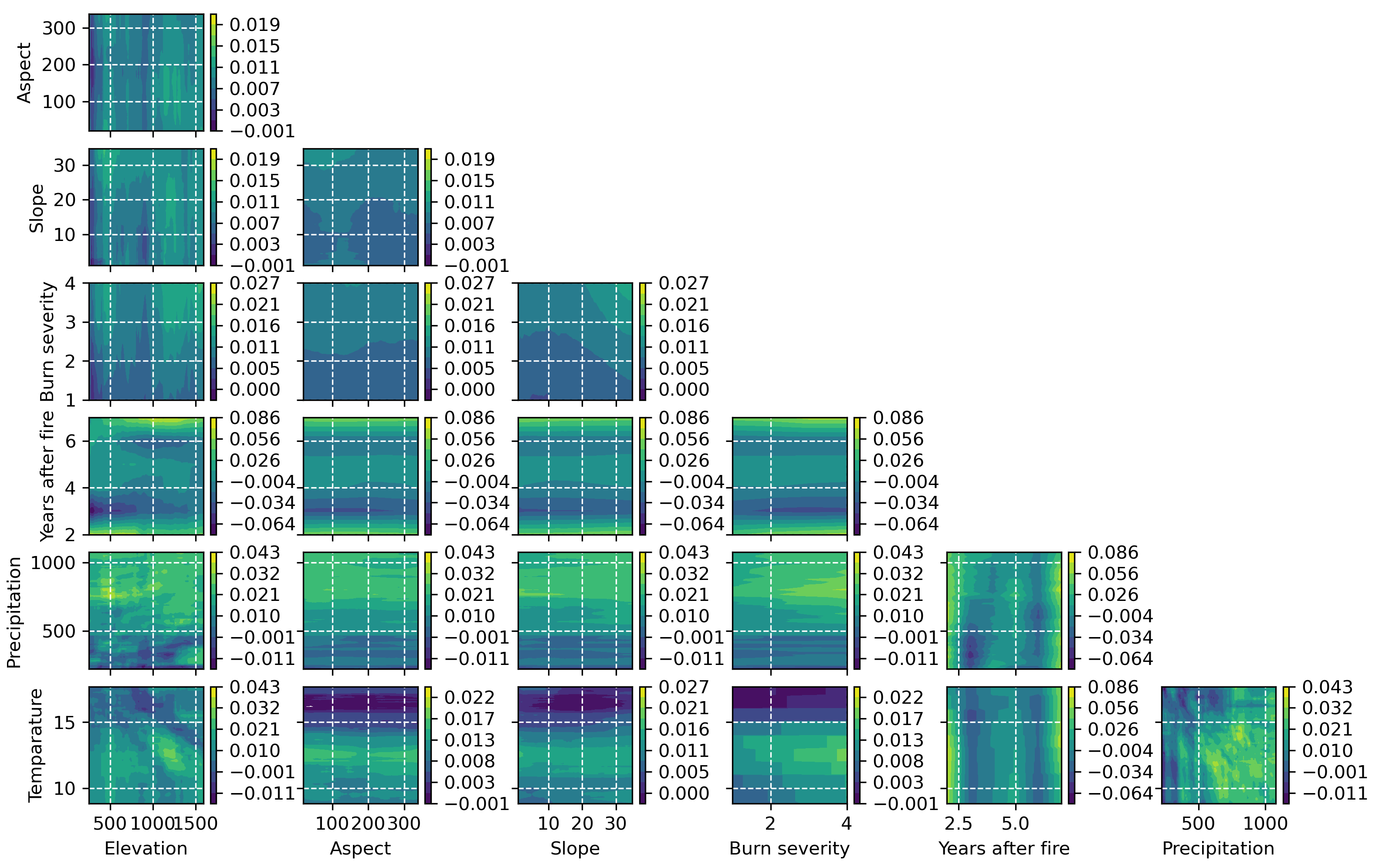}
    \caption{Two variable PDPs for GR showing the joint influence of variable pairs on predicting the target variable, incremental changes in EVI.}
    \label{fig:PDP_2var_GR}
\end{figure}

\begin{figure}[htbp]
    \centering
    \includegraphics[width=1\textwidth]{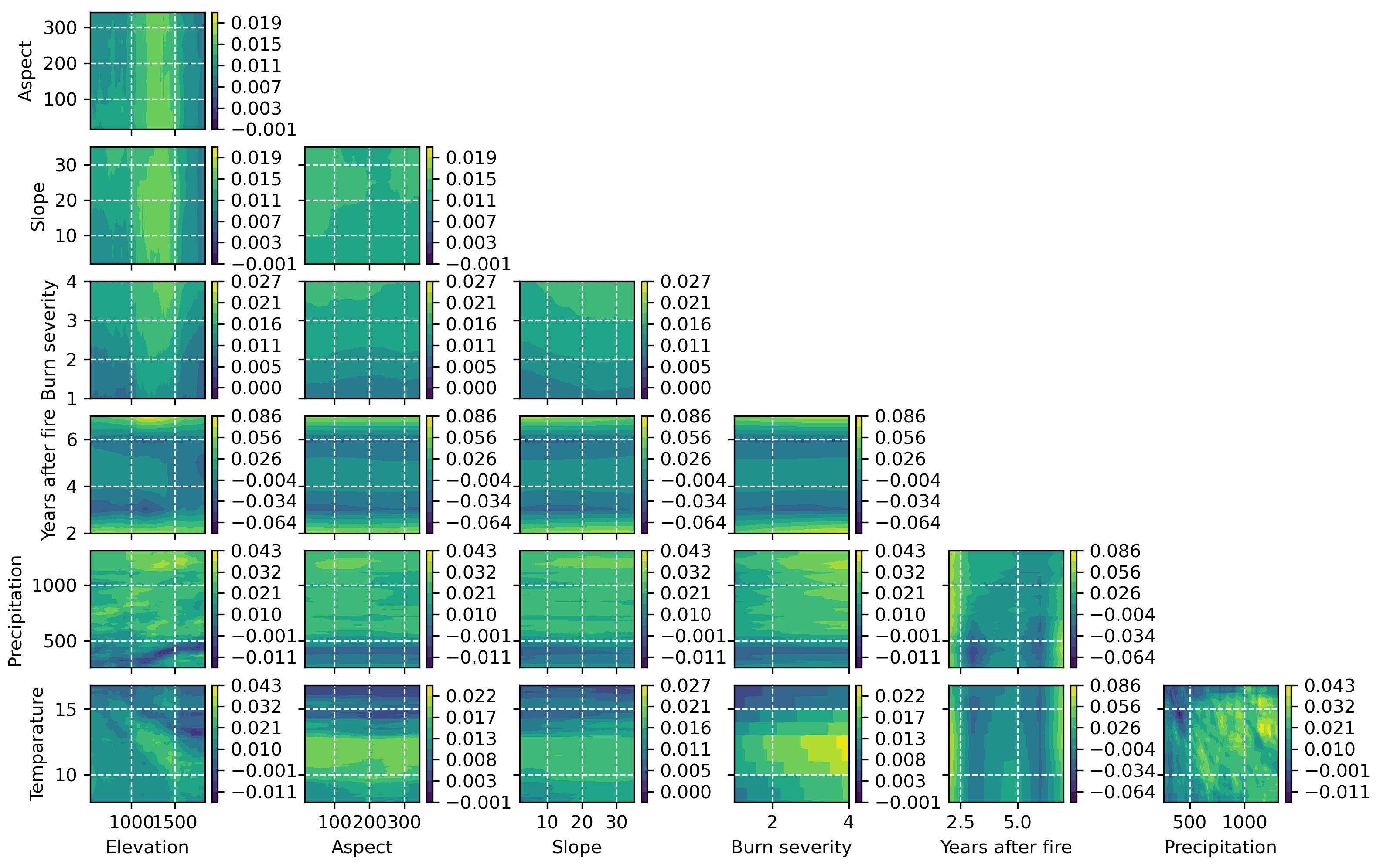}
    \caption{Two variable PDPs for SH showing the joint influence of variable pairs on predicting the target variable, incremental changes in EVI.}
    \label{fig:PDP_2var_SH}
\end{figure}

Two variable PDPs illustrate the interaction between features and their joint influence on incremental EVI changes. Two variable PDPs for EF, GR, and SH are shown in Figures~\ref{fig:PDP_2var_EF}, \ref{fig:PDP_2var_GR} and \ref{fig:PDP_2var_SH}, respectively.
Elevation-aspect pairs showed that lower elevation EF with east and south-facing aspects exhibit higher incremental EVI changes. In contrast, predictions for SH and GR showed changes with elevation while remaining unchanged along the aspect axis. Elevation-slope pairs showed that predicted PDP values remained flat for changing slope and varied predominantly along elevation for EF and SH. However, the plot for GR showed variation in incremental EVI changes along both axes. While values generally showed fluctuations, higher values appeared mostly on higher slopes and lower elevation. Slope-aspect pairs had the least variation in predicted incremental change in EVI values among all variable pairs.

Most of the pairs between burn severity and topographic features showed prediction changes mostly along burn severity and relatively less steep along topographic variables. Higher values occurred for higher burn severity and trends along the topographic features were consistent with the patterns observed in the one variable PDPs. Regarding the precipitation-burn severity pair, it was observed that low precipitation resulted in reduced PDP values, regardless of the level of burn severity. Generally, higher precipitation and burn severity led to higher values of incremental EVI changes. While Temperature-burn severity plots showed association of high temperature and high burn severity with higher PDP values for EF, SH and GR were found to have higher values within temperature ranges between 10 to 14\textdegree C.  

The years after the fire variable primarily influenced the PDP values for pairs where it was involved. When combining years after fire with topographic variables, burn severity, and temperature, there was significant variation in PDP values along the years after fire axis, while the other variable showed a minimal change in values. Higher PDP values occurred in the 2nd and 7th years after fire, while the 3rd and 6th year had the least recovery. Regarding years after fire and precipitation pairs, it was found that low total water year precipitation, approximately below 700 mm, during the 3rd and 6th years after the fire resulted in negative incremental EVI values for all land cover types. This suggests vegetation mortality or a lack of significant recovery during those periods due to insufficient moisture. On the other hand, higher incremental changes in EVI were observed in the 2nd year after the fire, regardless of the precipitation values, indicating a period of more rapid recovery early on after the fire event.

The precipitation-elevation pairs did not show a clear trend in predicting incremental changes in EVI. However, EF areas with total precipitation between 700-1300 mm and elevations below 1500 m experienced higher incremental EVI changes. Conversely, total precipitation below 500 mm resulted in lower values of the target variable, indicating limited vegetation recovery. For SH and GR, predicted values increased with higher precipitation, underscoring the significance of this variable in influencing incremental EVI changes. Regarding precipitation-aspect and precipitation-slope pairs for all land cover types, PDP values changed along the precipitation axis, with minimal changes along the aspect and slope axes. This suggested that precipitation had a more pronounced impact on the target variable, while aspect and slope had relatively less influence on the incremental changes in EVI.

Temperature-elevation pairs showed that at lower median annual temperatures, PDP values were low along changes in elevation. Higher incremental EVI values were predicted for temperature above 12.5\textdegree C and elevations below 1500 m for EF. No clear relationship were identified for SH and GR. However, higher temperatures above 13\textdegree C tended to have lower incremental EVI changes. Temperature-slope plots for SH and GR showed that incremental EVI changes were high for temperature ranges between 11 to 13\textdegree C. Below 11\textdegree C, the predicted values increased with increasing slope. For EF, predicted values showed variation mostly along temperature axis and values increased with increasing temperature. Similarly, temperature-aspect plots for all land cover types showed variation in PDP values predominantly along varying temperature. EF showed increasing PDP values with increasing temperature and SH and GR showed higher PDP values within an optimal temperature range (11 to 13\textdegree C).  

Finally, the climate variable pair (precipitation-temperature) exhibited a wide range of variations in predicting incremental EVI changes. EF had lower incremental EVI changes, even becoming negative which indicates vegetation destruction, for precipitation lower than 600 mm and temperatures below 9\textdegree C. Both SH and GR showed higher incremental EVI values with increasing temperature and precipitation. While precipitation below 500 mm tended to decrease incremental EVI changes, negative values occurred above 14\textdegree C for SH, and below 13 and above 15\textdegree C temperature for GR, indicating adverse effects on vegetation recovery under these conditions. The negative incremental EVI changes due to climate conditions did not appear in the one variable PDP plots, which highlights the importance of investigating the joint influence of features using two variable PDP analysis.

\section{Conclusion}
This study utilized a machine learning technique, RF regression, to model the spatial and temporal dynamics of post-fire vegetation recovery in Washington, Oregon, and Idaho, states in the PNW region of the United States. The research investigated the interplay between topographic and climate factors in influencing yearly vegetation recovery across different land cover types and burn severity levels following wildfires. 

The RF model demonstrated excellent predictive performance, with high $R^2$ scores between observed and predicted incremental EVI changes for both the training (0.99) and testing sets (between 0.89 and 0.945). Additionally, NSE and KGE scores were used to evaluate the model-predicted and observed total EVI recovery during each post-fire year. The results indicated that the RF model accurately captured the essential patterns and trends in incremental EVI changes without overfitting.

The study identified that climate variables, precipitation and temperature, were the most influential in predicting post-fire vegetation recovery. Excluding these variables caused notable drops in training and test $R^2$ scores, indicating their importance in predicting incremental EVI changes. The influence of these variables in predicting incremental EVI changes varied among the land cover types. EF exhibited increasing incremental EVI changes with increasing temperature within specific precipitation ranges. For GR and SH, higher post-fire precipitation and temperature were associated with increased vegetation recovery. For all land cover types, low precipitation during certain years after fire resulted in negative incremental EVI values, signifying vegetation mortality. Conversely, higher precipitation generally contributed to higher incremental EVI changes.

Among the topographic factors, elevation was found to significantly influence vegetation recovery, particularly for EF and SH land cover types. Aspect showed an influence on EF and exhibited higher incremental EVI changes on east and south-facing aspects. For SH and GR, aspect had minimal impact on incremental EVI changes, indicating a more uniform response to varying aspect values. While drop one variable analysis identified slope to be one of the least important features, one and two variable PDPs found that GR tended to have higher incremental EVI changes on steeper slopes. The impact of slope was negligible on EF and SH land cover types.

Investigating different burn severity levels revealed that high burn severity areas experienced higher incremental EVI changes compared to areas with low burn severity. Years elapsed after fire also had notable influence on vegetation recovery, with early post-fire year showing more rapid recovery.

Overall, this study found that different land cover types (EF, SH, and GH) exhibited varying responses to climate and topographic factors, indicating that the resilience and recovery of vegetation depend on the specific land cover characteristics. The results of this study could provide valuable insight for post-fire restoration planning. Furthermore, the machine learning approach used in this research can serve as a valuable tool for analyzing large-scale ecological changes following wildfires and other disturbances. Further research using diverse geographic regions would enhance the generalizability of the findings. The findings from this study contribute to a broader understanding of post-fire vegetation recovery dynamics, effective ecological restoration, and sustainable management of diverse ecosystem after fire in changing climate conditions.

\section{Acknowledgement}
This research was supported by the U.S. Department of Energy (DOE), Office of Science (SC) Biological and Environmental Research (BER) program, as part of BER's Environmental System Science (ESS) program. This contribution originates from the River Corridor Scientific Focus Area (SFA) at Pacific Northwest National Laboratory (PNNL). PNNL is operated for DOE by Battelle Memorial Institute under contract DE-AC05-76RL01830. This paper describes objective technical results and analysis. Any subjective views or opinions that might be expressed in the paper do not necessarily represent the views of the U.S. Department of Energy or the United States Government.

This research used resources of the National Energy Research Scientific Computing Center, a DOE Office of Science User Facility supported by the Office of Science of the U.S. Department of Energy under Contract No. DE-AC02-05CH11231 using NERSC award
BER-ERCAP0023098.

\section{Research Data}
The data and scripts used to generate the results in this manuscript are available on ESS-DIVE repository \citep{Zahura2023}.


\printcredits

\bibliographystyle{cas-model2-names}

\bibliography{cas-refs}

\begin{thebibliography}{67}
\expandafter\ifx\csname natexlab\endcsname\relax\def\natexlab#1{#1}\fi
\providecommand{\url}[1]{\texttt{#1}}
\providecommand{\href}[2]{#2}
\providecommand{\path}[1]{#1}
\providecommand{\DOIprefix}{doi:}
\providecommand{\ArXivprefix}{arXiv:}
\providecommand{\URLprefix}{URL: }
\providecommand{\Pubmedprefix}{pmid:}
\providecommand{\doi}[1]{\href{http://dx.doi.org/#1}{\path{#1}}}
\providecommand{\Pubmed}[1]{\href{pmid:#1}{\path{#1}}}
\providecommand{\bibinfo}[2]{#2}
\ifx\xfnm\relax \def\xfnm[#1]{\unskip,\space#1}\fi
\bibitem[{Abatzoglou and Williams(2016)}]{abatzoglou2016impact}
\bibinfo{author}{Abatzoglou, J.T.}, \bibinfo{author}{Williams, A.P.},
  \bibinfo{year}{2016}.
\newblock \bibinfo{title}{Impact of anthropogenic climate change on wildfire
  across western us forests}.
\newblock \bibinfo{journal}{Proceedings of the National Academy of Sciences}
  \bibinfo{volume}{113}, \bibinfo{pages}{11770--11775}.
\bibitem[{Breiman(2001)}]{breiman2001random}
\bibinfo{author}{Breiman, L.}, \bibinfo{year}{2001}.
\newblock \bibinfo{title}{Random forests}.
\newblock \bibinfo{journal}{Machine learning} \bibinfo{volume}{45},
  \bibinfo{pages}{5--32}.
\bibitem[{Casady et~al.(2010)Casady, van Leeuwen and Marsh}]{Casady2010}
\bibinfo{author}{Casady, G.M.}, \bibinfo{author}{van Leeuwen, W.J.},
  \bibinfo{author}{Marsh, S.E.}, \bibinfo{year}{2010}.
\newblock \bibinfo{title}{Evaluating post-wildfire vegetation regeneration as a
  response to multiple environmental determinants}.
\newblock \bibinfo{journal}{Environmental Modeling and Assessment}
  \bibinfo{volume}{15}, \bibinfo{pages}{295--307}.
\newblock \DOIprefix\doi{10.1007/s10666-009-9210-x}.
\bibitem[{Chappell and Agee(1996)}]{chappell1996fire}
\bibinfo{author}{Chappell, C.B.}, \bibinfo{author}{Agee, J.K.},
  \bibinfo{year}{1996}.
\newblock \bibinfo{title}{Fire severity and tree seedling establishment in
  abies magnifica forests, southern cascades, oregon}.
\newblock \bibinfo{journal}{Ecological Applications} \bibinfo{volume}{6},
  \bibinfo{pages}{628--640}.
\bibitem[{Chen and Chang(2022)}]{chen2022review}
\bibinfo{author}{Chen, J.}, \bibinfo{author}{Chang, H.}, \bibinfo{year}{2022}.
\newblock \bibinfo{title}{A review of wildfire impacts on stream temperature
  and turbidity across scales}.
\newblock \bibinfo{journal}{Progress in Physical Geography: Earth and
  Environment} , \bibinfo{pages}{03091333221118363}.
\bibitem[{Chen et~al.(2022)Chen, Chen and Xu}]{chen2022remote}
\bibinfo{author}{Chen, X.}, \bibinfo{author}{Chen, W.}, \bibinfo{author}{Xu,
  M.}, \bibinfo{year}{2022}.
\newblock \bibinfo{title}{Remote-sensing monitoring of postfire vegetation
  dynamics in the greater hinggan mountain range based on long time-series
  data: Analysis of the effects of six topographic and climatic factors}.
\newblock \bibinfo{journal}{Remote Sensing} \bibinfo{volume}{14},
  \bibinfo{pages}{2958}.
\bibitem[{Chen et~al.(2005)Chen, Vierling, Deering and
  Conley}]{chen2005monitoring}
\bibinfo{author}{Chen, X.}, \bibinfo{author}{Vierling, L.},
  \bibinfo{author}{Deering, D.}, \bibinfo{author}{Conley, A.},
  \bibinfo{year}{2005}.
\newblock \bibinfo{title}{Monitoring boreal forest leaf area index across a
  siberian burn chronosequence: a modis validation study}.
\newblock \bibinfo{journal}{International Journal of Remote Sensing}
  \bibinfo{volume}{26}, \bibinfo{pages}{5433--5451}.
\bibitem[{Cook et~al.(2018)Cook, Mankin and Anchukaitis}]{cook2018climate}
\bibinfo{author}{Cook, B.I.}, \bibinfo{author}{Mankin, J.S.},
  \bibinfo{author}{Anchukaitis, K.J.}, \bibinfo{year}{2018}.
\newblock \bibinfo{title}{Climate change and drought: From past to future}.
\newblock \bibinfo{journal}{Current Climate Change Reports}
  \bibinfo{volume}{4}, \bibinfo{pages}{164--179}.
\bibitem[{Daskalakou and Thanos(1996)}]{daskalakou1996aleppo}
\bibinfo{author}{Daskalakou, E.N.}, \bibinfo{author}{Thanos, C.A.},
  \bibinfo{year}{1996}.
\newblock \bibinfo{title}{Aleppo pine (pinus halepensis) postfire regeneration:
  the role of canopy and soil seed banks}.
\newblock \bibinfo{journal}{International Journal of Wildland Fire}
  \bibinfo{volume}{6}, \bibinfo{pages}{59--66}.
\bibitem[{Dewitz and USGS(2021)}]{NLCD2019}
\bibinfo{author}{Dewitz, J.}, \bibinfo{author}{USGS}, \bibinfo{year}{2021}.
\newblock \bibinfo{title}{National land cover database (nlcd) 2019 products
  (ver. 2.0, june 2021): U.s. geological survey data release}.
\newblock \URLprefix
  \url{https://developers.google.com/earth-engine/datasets/catalog/USGS_NLCD_RELEASES_2019_REL_NLCD#citations},
  \DOIprefix\doi{10.5066/P9KZCM54}.
\bibitem[{D{\'\i}az-Delgado et~al.(2002)D{\'\i}az-Delgado, Lloret, Pons and
  Terradas}]{diaz2002satellite}
\bibinfo{author}{D{\'\i}az-Delgado, R.}, \bibinfo{author}{Lloret, F.},
  \bibinfo{author}{Pons, X.}, \bibinfo{author}{Terradas, J.},
  \bibinfo{year}{2002}.
\newblock \bibinfo{title}{Satellite evidence of decreasing resilience in
  mediterranean plant communities after recurrent wildfires}.
\newblock \bibinfo{journal}{Ecology} \bibinfo{volume}{83},
  \bibinfo{pages}{2293--2303}.
\bibitem[{Didan(2021)}]{MOD13Q12021}
\bibinfo{author}{Didan, K.}, \bibinfo{year}{2021}.
\newblock \bibinfo{title}{Mod13q1 modis/terra vegetation indices 16-day l3
  global 250m sin grid v061}.
\bibitem[{Eidenshink et~al.(2007)Eidenshink, Schwind, Brewer, Zhu, Quayle and
  Howard}]{eidenshink2007project}
\bibinfo{author}{Eidenshink, J.}, \bibinfo{author}{Schwind, B.},
  \bibinfo{author}{Brewer, K.}, \bibinfo{author}{Zhu, Z.L.},
  \bibinfo{author}{Quayle, B.}, \bibinfo{author}{Howard, S.},
  \bibinfo{year}{2007}.
\newblock \bibinfo{title}{A project for monitoring trends in burn severity}.
\newblock \bibinfo{journal}{Fire ecology} \bibinfo{volume}{3},
  \bibinfo{pages}{3--21}.
\bibitem[{Evangelides and Nobajas(2020)}]{Evangelides2020}
\bibinfo{author}{Evangelides, C.}, \bibinfo{author}{Nobajas, A.},
  \bibinfo{year}{2020}.
\newblock \bibinfo{title}{Red-edge normalised difference vegetation index
  (ndvi705) from sentinel-2 imagery to assess post-fire regeneration}.
\newblock \bibinfo{journal}{Remote Sensing Applications: Society and
  Environment} \bibinfo{volume}{17}, \bibinfo{pages}{100283}.
\newblock \DOIprefix\doi{10.1016/J.RSASE.2019.100283}.
\bibitem[{Fox et~al.(2008)Fox, Maselli and Carrega}]{fox2008using}
\bibinfo{author}{Fox, D.}, \bibinfo{author}{Maselli, F.},
  \bibinfo{author}{Carrega, P.}, \bibinfo{year}{2008}.
\newblock \bibinfo{title}{Using spot images and field sampling to map burn
  severity and vegetation factors affecting post forest fire erosion risk}.
\newblock \bibinfo{journal}{Catena} \bibinfo{volume}{75},
  \bibinfo{pages}{326--335}.
\bibitem[{Friedman(2001)}]{friedman2001greedy}
\bibinfo{author}{Friedman, J.H.}, \bibinfo{year}{2001}.
\newblock \bibinfo{title}{Greedy function approximation: a gradient boosting
  machine}.
\newblock \bibinfo{journal}{Annals of statistics} ,
  \bibinfo{pages}{1189--1232}.
\bibitem[{Fultz et~al.(2016)Fultz, Moore-Kucera, Dathe, Davinic, Perry, Wester,
  Schwilk and Rideout-Hanzak}]{fultz2016forest}
\bibinfo{author}{Fultz, L.M.}, \bibinfo{author}{Moore-Kucera, J.},
  \bibinfo{author}{Dathe, J.}, \bibinfo{author}{Davinic, M.},
  \bibinfo{author}{Perry, G.}, \bibinfo{author}{Wester, D.},
  \bibinfo{author}{Schwilk, D.W.}, \bibinfo{author}{Rideout-Hanzak, S.},
  \bibinfo{year}{2016}.
\newblock \bibinfo{title}{Forest wildfire and grassland prescribed fire effects
  on soil biogeochemical processes and microbial communities: Two case studies
  in the semi-arid southwest}.
\newblock \bibinfo{journal}{Applied Soil Ecology} \bibinfo{volume}{99},
  \bibinfo{pages}{118--128}.
\bibitem[{Gao et~al.(2000)Gao, Huete, Ni and Miura}]{gao2000optical}
\bibinfo{author}{Gao, X.}, \bibinfo{author}{Huete, A.R.}, \bibinfo{author}{Ni,
  W.}, \bibinfo{author}{Miura, T.}, \bibinfo{year}{2000}.
\newblock \bibinfo{title}{Optical--biophysical relationships of vegetation
  spectra without background contamination}.
\newblock \bibinfo{journal}{Remote sensing of environment}
  \bibinfo{volume}{74}, \bibinfo{pages}{609--620}.
\bibitem[{Gergel et~al.(2017)Gergel, Nijssen, Abatzoglou, Lettenmaier and
  Stumbaugh}]{gergel2017effects}
\bibinfo{author}{Gergel, D.R.}, \bibinfo{author}{Nijssen, B.},
  \bibinfo{author}{Abatzoglou, J.T.}, \bibinfo{author}{Lettenmaier, D.P.},
  \bibinfo{author}{Stumbaugh, M.R.}, \bibinfo{year}{2017}.
\newblock \bibinfo{title}{Effects of climate change on snowpack and fire
  potential in the western usa}.
\newblock \bibinfo{journal}{Climatic Change} \bibinfo{volume}{141},
  \bibinfo{pages}{287--299}.
\bibitem[{Goetz et~al.(2006)Goetz, Fiske and Bunn}]{goetz2006using}
\bibinfo{author}{Goetz, S.J.}, \bibinfo{author}{Fiske, G.J.},
  \bibinfo{author}{Bunn, A.G.}, \bibinfo{year}{2006}.
\newblock \bibinfo{title}{Using satellite time-series data sets to analyze fire
  disturbance and forest recovery across canada}.
\newblock \bibinfo{journal}{Remote Sensing of Environment}
  \bibinfo{volume}{101}, \bibinfo{pages}{352--365}.
\bibitem[{Griffiths et~al.(2009)Griffiths, Madritch and
  Swanson}]{griffiths2009effects}
\bibinfo{author}{Griffiths, R.}, \bibinfo{author}{Madritch, M.},
  \bibinfo{author}{Swanson, A.}, \bibinfo{year}{2009}.
\newblock \bibinfo{title}{The effects of topography on forest soil
  characteristics in the oregon cascade mountains (usa): Implications for the
  effects of climate change on soil properties}.
\newblock \bibinfo{journal}{Forest Ecology and Management}
  \bibinfo{volume}{257}, \bibinfo{pages}{1--7}.
\bibitem[{Haffey et~al.(2018)Haffey, Sisk, Allen, Thode and
  Margolis}]{haffey2018limits}
\bibinfo{author}{Haffey, C.}, \bibinfo{author}{Sisk, T.D.},
  \bibinfo{author}{Allen, C.D.}, \bibinfo{author}{Thode, A.E.},
  \bibinfo{author}{Margolis, E.Q.}, \bibinfo{year}{2018}.
\newblock \bibinfo{title}{Limits to ponderosa pine regeneration following large
  high-severity forest fires in the united states southwest}.
\newblock \bibinfo{journal}{Fire Ecology} \bibinfo{volume}{14},
  \bibinfo{pages}{143--163}.
\bibitem[{Halofsky et~al.(2020)Halofsky, Peterson and
  Harvey}]{halofsky2020changing}
\bibinfo{author}{Halofsky, J.E.}, \bibinfo{author}{Peterson, D.L.},
  \bibinfo{author}{Harvey, B.J.}, \bibinfo{year}{2020}.
\newblock \bibinfo{title}{Changing wildfire, changing forests: the effects of
  climate change on fire regimes and vegetation in the pacific northwest, usa}.
\newblock \bibinfo{journal}{Fire Ecology} \bibinfo{volume}{16},
  \bibinfo{pages}{1--26}.
\bibitem[{Hao et~al.(2022)Hao, Xu, Wu and Tan}]{hao2022long}
\bibinfo{author}{Hao, B.}, \bibinfo{author}{Xu, X.}, \bibinfo{author}{Wu, F.},
  \bibinfo{author}{Tan, L.}, \bibinfo{year}{2022}.
\newblock \bibinfo{title}{Long-term effects of fire severity and climatic
  factors on post-forest-fire vegetation recovery}.
\newblock \bibinfo{journal}{Forests} \bibinfo{volume}{13},
  \bibinfo{pages}{883}.
\bibitem[{Harvey et~al.(2016)Harvey, Donato and Turner}]{harvey2016high}
\bibinfo{author}{Harvey, B.J.}, \bibinfo{author}{Donato, D.C.},
  \bibinfo{author}{Turner, M.G.}, \bibinfo{year}{2016}.
\newblock \bibinfo{title}{High and dry: Post-fire tree seedling establishment
  in subalpine forests decreases with post-fire drought and large
  stand-replacing burn patches}.
\newblock \bibinfo{journal}{Global Ecology and Biogeography}
  \bibinfo{volume}{25}, \bibinfo{pages}{655--669}.
\bibitem[{Holben(1986)}]{holben1986characteristics}
\bibinfo{author}{Holben, B.N.}, \bibinfo{year}{1986}.
\newblock \bibinfo{title}{Characteristics of maximum-value composite images
  from temporal avhrr data}.
\newblock \bibinfo{journal}{International journal of remote sensing}
  \bibinfo{volume}{7}, \bibinfo{pages}{1417--1434}.
\bibitem[{Hrelja et~al.(2020)Hrelja, {\v{S}}estak and
  Bogunovi{\'c}}]{hrelja2020wildfire}
\bibinfo{author}{Hrelja, I.}, \bibinfo{author}{{\v{S}}estak, I.},
  \bibinfo{author}{Bogunovi{\'c}, I.}, \bibinfo{year}{2020}.
\newblock \bibinfo{title}{Wildfire impacts on soil physical and chemical
  properties-a short review of recent studies}.
\newblock \bibinfo{journal}{Agriculturae Conspectus Scientificus}
  \bibinfo{volume}{85}, \bibinfo{pages}{293--301}.
\bibitem[{Huete et~al.(1985)Huete, Jackson and Post}]{huete1985spectral}
\bibinfo{author}{Huete, A.R.}, \bibinfo{author}{Jackson, R.D.},
  \bibinfo{author}{Post, D.}, \bibinfo{year}{1985}.
\newblock \bibinfo{title}{Spectral response of a plant canopy with different
  soil backgrounds}.
\newblock \bibinfo{journal}{Remote sensing of environment}
  \bibinfo{volume}{17}, \bibinfo{pages}{37--53}.
\bibitem[{Ireland and Petropoulos(2015)}]{ireland2015exploring}
\bibinfo{author}{Ireland, G.}, \bibinfo{author}{Petropoulos, G.P.},
  \bibinfo{year}{2015}.
\newblock \bibinfo{title}{Exploring the relationships between post-fire
  vegetation regeneration dynamics, topography and burn severity: A case study
  from the montane cordillera ecozones of western canada}.
\newblock \bibinfo{journal}{Applied Geography} \bibinfo{volume}{56},
  \bibinfo{pages}{232--248}.
\bibitem[{Johnstone et~al.(2010)Johnstone, McIntire, Pedersen, King and
  Pisaric}]{johnstone2010sensitive}
\bibinfo{author}{Johnstone, J.F.}, \bibinfo{author}{McIntire, E.J.},
  \bibinfo{author}{Pedersen, E.J.}, \bibinfo{author}{King, G.},
  \bibinfo{author}{Pisaric, M.J.}, \bibinfo{year}{2010}.
\newblock \bibinfo{title}{A sensitive slope: estimating landscape patterns of
  forest resilience in a changing climate}.
\newblock \bibinfo{journal}{Ecosphere} \bibinfo{volume}{1},
  \bibinfo{pages}{1--21}.
\bibitem[{Kinoshita and Hogue(2011)}]{kinoshita2011spatial}
\bibinfo{author}{Kinoshita, A.M.}, \bibinfo{author}{Hogue, T.S.},
  \bibinfo{year}{2011}.
\newblock \bibinfo{title}{Spatial and temporal controls on post-fire hydrologic
  recovery in southern california watersheds}.
\newblock \bibinfo{journal}{Catena} \bibinfo{volume}{87},
  \bibinfo{pages}{240--252}.
\bibitem[{Kraaij et~al.(2018)Kraaij, Baard, Arndt, Vhengani and
  Van~Wilgen}]{kraaij2018assessment}
\bibinfo{author}{Kraaij, T.}, \bibinfo{author}{Baard, J.A.},
  \bibinfo{author}{Arndt, J.}, \bibinfo{author}{Vhengani, L.},
  \bibinfo{author}{Van~Wilgen, B.W.}, \bibinfo{year}{2018}.
\newblock \bibinfo{title}{An assessment of climate, weather, and fuel factors
  influencing a large, destructive wildfire in the knysna region, south
  africa}.
\newblock \bibinfo{journal}{Fire Ecology} \bibinfo{volume}{14},
  \bibinfo{pages}{1--12}.
\bibitem[{scikit learn(2023a)}]{RanForReg2023}
\bibinfo{author}{scikit learn}, \bibinfo{year}{2023}a.
\newblock \bibinfo{title}{sklearn.ensemble.randomforestregressor —
  scikit-learn 1.3.0 documentation}.
\newblock
  \bibinfo{howpublished}{\url{https://scikit-learn.org/stable/modules/generated/sklearn.ensemble.RandomForestRegressor.html}}.
\bibitem[{scikit learn(2023b)}]{PartialDependence_2023}
\bibinfo{author}{scikit learn}, \bibinfo{year}{2023}b.
\newblock \bibinfo{title}{sklearn.inspection.partial\_dependence —
  scikit-learn 1.3.0 documentation}.
\newblock
  \bibinfo{howpublished}{\url{https://scikit-learn.org/stable/modules/generated/sklearn.inspection.partial_dependence.html}}.
\bibitem[{scikit learn(2023c)}]{GridSearchCV_2023}
\bibinfo{author}{scikit learn}, \bibinfo{year}{2023}c.
\newblock \bibinfo{title}{sklearn.model\_selection.gridsearchcv —
  scikit-learn 1.3.0 documentation}.
\newblock
  \bibinfo{howpublished}{\url{https://scikit-learn.org/stable/modules/generated/sklearn.model_selection.GridSearchCV.html}}.
\bibitem[{Lentile et~al.(2006)Lentile, Holden, Smith, Falkowski, Hudak, Morgan,
  Lewis, Gessler and Benson}]{lentile2006remote}
\bibinfo{author}{Lentile, L.B.}, \bibinfo{author}{Holden, Z.A.},
  \bibinfo{author}{Smith, A.M.}, \bibinfo{author}{Falkowski, M.J.},
  \bibinfo{author}{Hudak, A.T.}, \bibinfo{author}{Morgan, P.},
  \bibinfo{author}{Lewis, S.A.}, \bibinfo{author}{Gessler, P.E.},
  \bibinfo{author}{Benson, N.C.}, \bibinfo{year}{2006}.
\newblock \bibinfo{title}{Remote sensing techniques to assess active fire
  characteristics and post-fire effects}.
\newblock \bibinfo{journal}{International Journal of Wildland Fire}
  \bibinfo{volume}{15}, \bibinfo{pages}{319--345}.
\bibitem[{Leon et~al.(2012)Leon, Van~Leeuwen and Casady}]{leon2012using}
\bibinfo{author}{Leon, J.R.R.}, \bibinfo{author}{Van~Leeuwen, W.J.},
  \bibinfo{author}{Casady, G.M.}, \bibinfo{year}{2012}.
\newblock \bibinfo{title}{Using modis-ndvi for the modeling of post-wildfire
  vegetation response as a function of environmental conditions and pre-fire
  restoration treatments}.
\newblock \bibinfo{journal}{Remote sensing} \bibinfo{volume}{4},
  \bibinfo{pages}{598--621}.
\bibitem[{Lippok et~al.(2013)Lippok, Beck, Renison, Gallegos, Saavedra, Hensen
  and Schleuning}]{lippok2013forest}
\bibinfo{author}{Lippok, D.}, \bibinfo{author}{Beck, S.G.},
  \bibinfo{author}{Renison, D.}, \bibinfo{author}{Gallegos, S.C.},
  \bibinfo{author}{Saavedra, F.V.}, \bibinfo{author}{Hensen, I.},
  \bibinfo{author}{Schleuning, M.}, \bibinfo{year}{2013}.
\newblock \bibinfo{title}{Forest recovery of areas deforested by fire increases
  with elevation in the tropical andes}.
\newblock \bibinfo{journal}{Forest Ecology and Management}
  \bibinfo{volume}{295}, \bibinfo{pages}{69--76}.
\bibitem[{Liu and Huete(1995)}]{liu1995feedback}
\bibinfo{author}{Liu, H.Q.}, \bibinfo{author}{Huete, A.}, \bibinfo{year}{1995}.
\newblock \bibinfo{title}{A feedback based modification of the ndvi to minimize
  canopy background and atmospheric noise}.
\newblock \bibinfo{journal}{IEEE transactions on geoscience and remote sensing}
  \bibinfo{volume}{33}, \bibinfo{pages}{457--465}.
\bibitem[{Maia et~al.(2014)Maia, Keizer, Vasques, Abrantes, Roxo, Fernandes,
  Ferreira and Moreira}]{Maia2014}
\bibinfo{author}{Maia, P.}, \bibinfo{author}{Keizer, J.},
  \bibinfo{author}{Vasques, A.}, \bibinfo{author}{Abrantes, N.},
  \bibinfo{author}{Roxo, L.}, \bibinfo{author}{Fernandes, P.},
  \bibinfo{author}{Ferreira, A.}, \bibinfo{author}{Moreira, F.},
  \bibinfo{year}{2014}.
\newblock \bibinfo{title}{Post-fire plant diversity and abundance in pine and
  eucalypt stands in portugal: Effects of biogeography, topography, forest type
  and post-fire management}.
\newblock \bibinfo{journal}{Forest Ecology and Management}
  \bibinfo{volume}{334}, \bibinfo{pages}{154--162}.
\newblock \DOIprefix\doi{10.1016/j.foreco.2014.08.030}.
\bibitem[{Malak and Pausas(2006)}]{malak2006fire}
\bibinfo{author}{Malak, D.A.}, \bibinfo{author}{Pausas, J.G.},
  \bibinfo{year}{2006}.
\newblock \bibinfo{title}{Fire regime and post-fire normalized difference
  vegetation index changes in the eastern iberian peninsula (mediterranean
  basin)}.
\newblock \bibinfo{journal}{International Journal of Wildland Fire}
  \bibinfo{volume}{15}, \bibinfo{pages}{407--413}.
\bibitem[{Marlier et~al.(2017)Marlier, Xiao, Engel, Livneh, Abatzoglou and
  Lettenmaier}]{marlier20172015}
\bibinfo{author}{Marlier, M.E.}, \bibinfo{author}{Xiao, M.},
  \bibinfo{author}{Engel, R.}, \bibinfo{author}{Livneh, B.},
  \bibinfo{author}{Abatzoglou, J.T.}, \bibinfo{author}{Lettenmaier, D.P.},
  \bibinfo{year}{2017}.
\newblock \bibinfo{title}{The 2015 drought in washington state: a harbinger of
  things to come?}
\newblock \bibinfo{journal}{Environmental Research Letters}
  \bibinfo{volume}{12}, \bibinfo{pages}{114008}.
\bibitem[{Meng et~al.(2015)Meng, Dennison, Huang, Moritz and
  D'Antonio}]{meng2015effects}
\bibinfo{author}{Meng, R.}, \bibinfo{author}{Dennison, P.E.},
  \bibinfo{author}{Huang, C.}, \bibinfo{author}{Moritz, M.A.},
  \bibinfo{author}{D'Antonio, C.}, \bibinfo{year}{2015}.
\newblock \bibinfo{title}{Effects of fire severity and post-fire climate on
  short-term vegetation recovery of mixed-conifer and red fir forests in the
  sierra nevada mountains of california}.
\newblock \bibinfo{journal}{Remote Sensing of Environment}
  \bibinfo{volume}{171}, \bibinfo{pages}{311--325}.
\bibitem[{Mote et~al.(2016)Mote, Rupp, Li, Sharp, Otto, Uhe, Xiao, Lettenmaier,
  Cullen and Allen}]{mote2016perspectives}
\bibinfo{author}{Mote, P.W.}, \bibinfo{author}{Rupp, D.E.},
  \bibinfo{author}{Li, S.}, \bibinfo{author}{Sharp, D.J.},
  \bibinfo{author}{Otto, F.}, \bibinfo{author}{Uhe, P.F.},
  \bibinfo{author}{Xiao, M.}, \bibinfo{author}{Lettenmaier, D.P.},
  \bibinfo{author}{Cullen, H.}, \bibinfo{author}{Allen, M.R.},
  \bibinfo{year}{2016}.
\newblock \bibinfo{title}{Perspectives on the causes of exceptionally low 2015
  snowpack in the western united states}.
\newblock \bibinfo{journal}{Geophysical Research Letters} \bibinfo{volume}{43},
  \bibinfo{pages}{10--980}.
\bibitem[{{MTBS Project (USDA Forest Service/U.S. Geological
  Survey)}(2023)}]{MTBS2022}
\bibinfo{author}{{MTBS Project (USDA Forest Service/U.S. Geological Survey)}},
  \bibinfo{year}{2023}.
\newblock \bibinfo{title}{Mtbs data access: Fire level geospatial data}.
\newblock \URLprefix \url{https://www.mtbs.gov/direct-download}.
\bibitem[{Nelson et~al.(2013)Nelson, Weisberg and
  Kitchen}]{nelson2013influence}
\bibinfo{author}{Nelson, Z.J.}, \bibinfo{author}{Weisberg, P.J.},
  \bibinfo{author}{Kitchen, S.G.}, \bibinfo{year}{2013}.
\newblock \bibinfo{title}{Influence of climate and environment on post-fire
  recovery of mountain big sagebrush}.
\newblock \bibinfo{journal}{International Journal of Wildland Fire}
  \bibinfo{volume}{23}, \bibinfo{pages}{131--142}.
\bibitem[{Pedregosa et~al.(2011)Pedregosa, Varoquaux, Gramfort, Michel,
  Thirion, Grisel, Blondel, Prettenhofer, Weiss, Dubourg, Vanderplas, Passos,
  Cournapeau, Brucher, Perrot and Duchesnay}]{scikitlearn}
\bibinfo{author}{Pedregosa, F.}, \bibinfo{author}{Varoquaux, G.},
  \bibinfo{author}{Gramfort, A.}, \bibinfo{author}{Michel, V.},
  \bibinfo{author}{Thirion, B.}, \bibinfo{author}{Grisel, O.},
  \bibinfo{author}{Blondel, M.}, \bibinfo{author}{Prettenhofer, P.},
  \bibinfo{author}{Weiss, R.}, \bibinfo{author}{Dubourg, V.},
  \bibinfo{author}{Vanderplas, J.}, \bibinfo{author}{Passos, A.},
  \bibinfo{author}{Cournapeau, D.}, \bibinfo{author}{Brucher, M.},
  \bibinfo{author}{Perrot, M.}, \bibinfo{author}{Duchesnay, E.},
  \bibinfo{year}{2011}.
\newblock \bibinfo{title}{Scikit-learn: Machine learning in {P}ython}.
\newblock \bibinfo{journal}{Journal of Machine Learning Research}
  \bibinfo{volume}{12}, \bibinfo{pages}{2825--2830}.
\bibitem[{Peltier and Ogle(2019)}]{peltier2019legacies}
\bibinfo{author}{Peltier, D.M.}, \bibinfo{author}{Ogle, K.},
  \bibinfo{year}{2019}.
\newblock \bibinfo{title}{Legacies of more frequent drought in ponderosa pine
  across the western united states}.
\newblock \bibinfo{journal}{Global change biology} \bibinfo{volume}{25},
  \bibinfo{pages}{3803--3816}.
\bibitem[{Pereira et~al.(2016)Pereira, Cerd{\`a}, Lopez, Zavala, Mataix-Solera,
  Arcenegui, Misiune, Keesstra and Novara}]{pereira2016short}
\bibinfo{author}{Pereira, P.}, \bibinfo{author}{Cerd{\`a}, A.},
  \bibinfo{author}{Lopez, A.J.}, \bibinfo{author}{Zavala, L.M.},
  \bibinfo{author}{Mataix-Solera, J.}, \bibinfo{author}{Arcenegui, V.},
  \bibinfo{author}{Misiune, I.}, \bibinfo{author}{Keesstra, S.},
  \bibinfo{author}{Novara, A.}, \bibinfo{year}{2016}.
\newblock \bibinfo{title}{Short-term vegetation recovery after a grassland fire
  in lithuania: The effects of fire severity, slope position and aspect}.
\newblock \bibinfo{journal}{Land Degradation \& Development}
  \bibinfo{volume}{27}, \bibinfo{pages}{1523--1534}.
\bibitem[{Petropoulos et~al.(2014)Petropoulos, Griffiths and
  Kalivas}]{petropoulos2014quantifying}
\bibinfo{author}{Petropoulos, G.P.}, \bibinfo{author}{Griffiths, H.M.},
  \bibinfo{author}{Kalivas, D.P.}, \bibinfo{year}{2014}.
\newblock \bibinfo{title}{Quantifying spatial and temporal vegetation recovery
  dynamics following a wildfire event in a mediterranean landscape using eo
  data and gis}.
\newblock \bibinfo{journal}{Applied Geography} \bibinfo{volume}{50},
  \bibinfo{pages}{120--131}.
\bibitem[{Prodon and Diaz-Delgado(2021)}]{prodon2021assessing}
\bibinfo{author}{Prodon, R.}, \bibinfo{author}{Diaz-Delgado, R.},
  \bibinfo{year}{2021}.
\newblock \bibinfo{title}{Assessing the postfire resilience of a mediterranean
  forest from satellite and ground data (ndvi, vegetation profile, avifauna)}.
\newblock \bibinfo{journal}{{\'E}coscience} \bibinfo{volume}{28},
  \bibinfo{pages}{81--91}.
\bibitem[{{QGIS Development Team}(2023)}]{QGIS_software}
\bibinfo{author}{{QGIS Development Team}}, \bibinfo{year}{2023}.
\newblock \bibinfo{title}{QGIS Geographic Information System}.
\newblock \bibinfo{organization}{QGIS Association}.
\newblock \URLprefix \url{https://www.qgis.org}.
\bibitem[{Rengers et~al.(2020)Rengers, McGuire, Oakley, Kean, Staley and
  Tang}]{rengers2020landslides}
\bibinfo{author}{Rengers, F.K.}, \bibinfo{author}{McGuire, L.A.},
  \bibinfo{author}{Oakley, N.S.}, \bibinfo{author}{Kean, J.W.},
  \bibinfo{author}{Staley, D.M.}, \bibinfo{author}{Tang, H.},
  \bibinfo{year}{2020}.
\newblock \bibinfo{title}{Landslides after wildfire: Initiation, magnitude, and
  mobility}.
\newblock \bibinfo{journal}{Landslides} \bibinfo{volume}{17},
  \bibinfo{pages}{2631--2641}.
\bibitem[{Robinne et~al.(2020)Robinne, Hallema, Bladon and
  Buttle}]{robinne2020wildfire}
\bibinfo{author}{Robinne, F.N.}, \bibinfo{author}{Hallema, D.W.},
  \bibinfo{author}{Bladon, K.D.}, \bibinfo{author}{Buttle, J.M.},
  \bibinfo{year}{2020}.
\newblock \bibinfo{title}{Wildfire impacts on hydrologic ecosystem services in
  north american high-latitude forests: A scoping review}.
\newblock \bibinfo{journal}{Journal of Hydrology} \bibinfo{volume}{581},
  \bibinfo{pages}{124360}.
\bibitem[{Rouse~Jr et~al.(1974)Rouse~Jr, Haas, Deering, Schell and
  Harlan}]{rouse1974monitoring}
\bibinfo{author}{Rouse~Jr, J.W.}, \bibinfo{author}{Haas, R.H.},
  \bibinfo{author}{Deering, D.}, \bibinfo{author}{Schell, J.},
  \bibinfo{author}{Harlan, J.C.}, \bibinfo{year}{1974}.
\newblock \bibinfo{title}{Monitoring the vernal advancement and retrogradation
  (green wave effect) of natural vegetation}.
\newblock \bibinfo{type}{Technical Report}.
\bibitem[{S{\'a}nchez-G{\'o}mez et~al.(2006)S{\'a}nchez-G{\'o}mez, Valladares
  and Zavala}]{sanchez2006performance}
\bibinfo{author}{S{\'a}nchez-G{\'o}mez, D.}, \bibinfo{author}{Valladares, F.},
  \bibinfo{author}{Zavala, M.A.}, \bibinfo{year}{2006}.
\newblock \bibinfo{title}{Performance of seedlings of mediterranean woody
  species under experimental gradients of irradiance and water availability:
  trade-offs and evidence for niche differentiation}.
\newblock \bibinfo{journal}{New phytologist} \bibinfo{volume}{170},
  \bibinfo{pages}{795--806}.
\bibitem[{Taylor et~al.(2021)Taylor, Poulos, Kluber, Issacs, Pawlikowski and
  Barton}]{taylor2021controls}
\bibinfo{author}{Taylor, A.H.}, \bibinfo{author}{Poulos, H.M.},
  \bibinfo{author}{Kluber, J.}, \bibinfo{author}{Issacs, R.},
  \bibinfo{author}{Pawlikowski, N.}, \bibinfo{author}{Barton, A.M.},
  \bibinfo{year}{2021}.
\newblock \bibinfo{title}{Controls on spatial patterns of wildfire severity and
  early post-fire vegetation development in an arizona sky island, usa}.
\newblock \bibinfo{journal}{Landscape Ecology} \bibinfo{volume}{36},
  \bibinfo{pages}{2637--2656}.
\bibitem[{Thornton et~al.(2022)Thornton, Shrestha, Wei, Thornton, Kao and
  Wilson}]{https://doi.org/10.3334/ornldaac/2131}
\bibinfo{author}{Thornton, M.}, \bibinfo{author}{Shrestha, R.},
  \bibinfo{author}{Wei, Y.}, \bibinfo{author}{Thornton, P.},
  \bibinfo{author}{Kao, S.C.}, \bibinfo{author}{Wilson, B.},
  \bibinfo{year}{2022}.
\newblock \bibinfo{title}{Daymet: Monthly climate summaries on a 1-km grid for
  north america, version 4 r1}.
\newblock \URLprefix
  \url{https://daac.ornl.gov/cgi-bin/dsviewer.pl?ds_id=2131},
  \DOIprefix\doi{10.3334/ORNLDAAC/2131}.
\bibitem[{{U.S. Geological Survey}(2023)}]{USGS}
\bibinfo{author}{{U.S. Geological Survey}}, \bibinfo{year}{2023}.
\newblock \bibinfo{title}{3d elevation program 1-meter resolution digital
  elevation model v2}.
\newblock \URLprefix \url{https://apps.nationalmap.gov/downloader/}.
\bibitem[{Vallejo and Alloza(2012)}]{vallejo2012post}
\bibinfo{author}{Vallejo, V.R.}, \bibinfo{author}{Alloza, J.A.},
  \bibinfo{year}{2012}.
\newblock \bibinfo{title}{Post-fire management in the mediterranean basin}.
\newblock \bibinfo{journal}{Israel Journal of Ecology and Evolution}
  \bibinfo{volume}{58}, \bibinfo{pages}{251--264}.
\bibitem[{Viana-Soto et~al.(2017)Viana-Soto, Aguado and
  Mart{\'\i}nez}]{viana2017assessment}
\bibinfo{author}{Viana-Soto, A.}, \bibinfo{author}{Aguado, I.},
  \bibinfo{author}{Mart{\'\i}nez, S.}, \bibinfo{year}{2017}.
\newblock \bibinfo{title}{Assessment of post-fire vegetation recovery using
  fire severity and geographical data in the mediterranean region (spain)}.
\newblock \bibinfo{journal}{Environments} \bibinfo{volume}{4},
  \bibinfo{pages}{90}.
\bibitem[{Viana-Soto et~al.(2020)Viana-Soto, Aguado, Salas and
  Garc{\'\i}a}]{viana2020identifying}
\bibinfo{author}{Viana-Soto, A.}, \bibinfo{author}{Aguado, I.},
  \bibinfo{author}{Salas, J.}, \bibinfo{author}{Garc{\'\i}a, M.},
  \bibinfo{year}{2020}.
\newblock \bibinfo{title}{Identifying post-fire recovery trajectories and
  driving factors using landsat time series in fire-prone mediterranean pine
  forests}.
\newblock \bibinfo{journal}{Remote Sensing} \bibinfo{volume}{12},
  \bibinfo{pages}{1499}.
\bibitem[{Vo and Kinoshita(2020)}]{vo2020remote}
\bibinfo{author}{Vo, V.D.}, \bibinfo{author}{Kinoshita, A.M.},
  \bibinfo{year}{2020}.
\newblock \bibinfo{title}{Remote sensing of vegetation conditions after
  post-fire mulch treatments}.
\newblock \bibinfo{journal}{Journal of environmental management}
  \bibinfo{volume}{260}, \bibinfo{pages}{109993}.
\bibitem[{Wang et~al.(2022)Wang, Wang, Yong, Zhao, Xiong, Du and
  Zhao}]{wang2022vegetation}
\bibinfo{author}{Wang, Q.}, \bibinfo{author}{Wang, Z.g.},
  \bibinfo{author}{Yong, Z.w.}, \bibinfo{author}{Zhao, K.},
  \bibinfo{author}{Xiong, J.n.}, \bibinfo{author}{Du, X.m.},
  \bibinfo{author}{Zhao, Y.}, \bibinfo{year}{2022}.
\newblock \bibinfo{title}{Vegetation recovery trends under dual dominance of
  climate change and anthropogenic factors in the severely damaged areas of the
  wenchuan earthquake}.
\newblock \bibinfo{journal}{Journal of Mountain Science} \bibinfo{volume}{19},
  \bibinfo{pages}{3131--3147}.
\bibitem[{Wilson et~al.(2015)Wilson, Latimer and
  Silander~Jr}]{wilson2015climatic}
\bibinfo{author}{Wilson, A.M.}, \bibinfo{author}{Latimer, A.M.},
  \bibinfo{author}{Silander~Jr, J.A.}, \bibinfo{year}{2015}.
\newblock \bibinfo{title}{Climatic controls on ecosystem resilience: Postfire
  regeneration in the cape floristic region of south africa}.
\newblock \bibinfo{journal}{Proceedings of the National Academy of Sciences}
  \bibinfo{volume}{112}, \bibinfo{pages}{9058--9063}.
\bibitem[{Wittenberg et~al.(2007)Wittenberg, Malkinson, Beeri, Halutzy and
  Tesler}]{wittenberg2007spatial}
\bibinfo{author}{Wittenberg, L.}, \bibinfo{author}{Malkinson, D.},
  \bibinfo{author}{Beeri, O.}, \bibinfo{author}{Halutzy, A.},
  \bibinfo{author}{Tesler, N.}, \bibinfo{year}{2007}.
\newblock \bibinfo{title}{Spatial and temporal patterns of vegetation recovery
  following sequences of forest fires in a mediterranean landscape, mt. carmel
  israel}.
\newblock \bibinfo{journal}{Catena} \bibinfo{volume}{71},
  \bibinfo{pages}{76--83}.
\bibitem[{Zahura et~al.(2023)Zahura, Bisht, Li, McKnight and Chen}]{Zahura2023}
\bibinfo{author}{Zahura, F.}, \bibinfo{author}{Bisht, G.}, \bibinfo{author}{Li,
  Z.}, \bibinfo{author}{McKnight, S.}, \bibinfo{author}{Chen, X.},
  \bibinfo{year}{2023}.
\newblock \bibinfo{title}{Data, scripts, and figures associated with a
  manuscript studying impact of climate and topography on post-fire vegetation
  recovery}.
\newblock
  \bibinfo{howpublished}{\url{https://data.ess-dive.lbl.gov/datasets/doi:10.15485/2205677}}.
\newblock \bibinfo{note}{[accessed 2023-11-15]}.

\end{thebibliography}

\end{document}